\documentclass{article}

\usepackage{hyperref}

\usepackage{graphicx}
\usepackage{epstopdf}

\usepackage{amsmath,amssymb,amsthm} % AMS math packages
\usepackage{bm} % Bold in math mode
\usepackage{array} % for better arrays in maths
\usepackage{amscd} % for commutative diagrams

\newtheorem{Def}{Definition}[section]
\newtheorem{Lem}[Def]{Lemma}
\newtheorem{The}[Def]{Theorem}
\newtheorem{Cor}[Def]{Corollary}
\newtheorem{Prop}[Def]{Proposition}
\newtheorem{Exa}[Def]{Example}

\newcommand{\Tr}[1]{\ensuremath{\text{Tr} \left ( #1 \right )}}  % Trace
\newcommand{\PTr}[2]{\ensuremath{\text{Tr}_{#1} \left ( #2 \right )}}  % Partial Trace
\newcommand{\Ket}[1]{\ensuremath{\left | #1 \right \rangle}}  % Dirac Ket
\newcommand{\Bra}[1]{\ensuremath{\left \langle #1 \right |}}  % Dirac Bra
\def\kb#1#2{| #1 \rangle\!\langle #2 |}

\newcommand{\1}{\ensuremath{}}  
\newcommand{\ity}{\ensuremath{^{(\infty)}}}
\newcommand{\ns}[1]{\ensuremath{^{(#1)}}}
\newcommand{\pns}[1]{\ensuremath{\po}\ns{#1}}
\newcommand{\po}{\ensuremath{\star}}
\newcommand{\pity}{\ensuremath{\odot}}

\title{Quantum Graphical Models and Belief Propagation}
\author{M. S. Leifer\footnote{Institute for Quantum Computing, University of Waterloo, 200 University Ave. W., Waterloo, ON, Canada, N2L 3G1} $^,$\footnote{Perimeter Institute for Theoretical Physics, 31 Caroline St. N., Waterloo, ON, Canada, N2L 2Y5} $^,$\footnote{\tt{matt@mattleifer.info}} \and D. Poulin\footnote{Center for the Physics of Information, California Institute of Technology, Pasadena, CA 91125} $^,$\footnote{\tt{dpoulin@ist.caltech.edu}}}
\date{\today}                                          

\begin{document}
\maketitle

% ***** ABSTRACT *****

\begin{abstract}
Belief Propagation algorithms acting on Graphical Models of classical probability distributions, such as Markov Networks, Factor Graphs and Bayesian Networks, are amongst the most powerful known methods for deriving probabilistic inferences amongst large numbers of random variables.  This paper presents a generalization of these concepts and methods to the quantum case, based on the idea that quantum theory can be thought of as a noncommutative, operator-valued, generalization of classical probability theory.  Some novel characterizations of quantum conditional independence are derived, and definitions of Quantum $n$-Bifactor Networks, Markov Networks, Factor Graphs and Bayesian Networks are proposed.  The structure of Quantum Markov Networks is investigated and some partial characterization results are obtained, along the lines of the Hammersely-Clifford theorem.  A Quantum Belief Propagation algorithm is presented and is shown to converge on $1$-Bifactor Networks and Markov Networks when the underlying graph is a tree.  The use of Quantum Belief Propagation as a heuristic algorithm in cases where it is not known to converge is discussed.  Applications to decoding quantum error correcting codes and to the simulation of many-body quantum systems are described. 
\end{abstract}
%\keywords{quantum conditional probability, graphical models, Markov Networks, Belief Propagation, quantum error correction, quantum simulation}
%\pacs{XXXXX}

% ***** 1) INTRODUCTION *****

\section{Introduction}

\label{Intro}

Quantum theory is first and foremost a calculus for computing the probabilities of outcomes of measurements made on physical systems.  Therefore, the generic problem in quantum theory is one of probabilistic inference, i.e. given a specified class of quantum states, compute the predicted probabilities of measurement outcomes and their correlations.  For example, computing the correlation functions of a system in the ground state of a Hamiltonian, or computing the probabilities for the possible measurement outcomes after implementing a quantum circuit, are problems of this general type.  Such quantum inferences present a formidable computational challenge as the number of subsystems becomes large, since the number of parameters needed to specify a quantum state grows exponentially with the number of subsystems, and the formulas for quantities of interest typically also involve an exponentially large number of terms. 

A similar problem arises in classical probabilistic inference, since the number of terms required to specify a general probability distribution also grows exponentially with the number of random variables involved.  A variety of algorithms for classical probabilistic inference have been discovered, of which Belief Propagation algorithms on Graphical Models are amongst the most powerful.  Such algorithms are particularly interesting for two reasons.  Firstly, they are highly parallelizable in the sense that they can be implemented by associating each random variable with a separate processor.  Messages are received and sent by the processors along the links of a network corresponding to the edges of a graph and, importantly, the order in which the messages arrive does not matter.  Secondly, Belief Propagation performs remarkably well as a heuristic algorithm, even in cases where it is not guaranteed to converge to the exact solution.  Important examples include the near optimal decoding of low density \cite{Gal63a} and turbo \cite{BGT93a} error correction codes, spin glass models \cite{MP01a}, and random satisfiability problems \cite{MPZ02a}. Understanding the reasons for this is currently an active area of research, but it is understood \cite{Yed01a} to be related to a hierarchy of approximation schemes commonly used in statistical physics.  

Due to the similarity between the classical and quantum problems, one might hope to leverage the power of Belief Propagation in the quantum case also, especially since quantum theory can be regarded as a noncommutative generalization of classical probability theory.  This is indeed the case, and in this paper we develop the necessary theory of Quantum Belief Propagation and its associated Graphical Models.

This paper should be of interest to researchers in Graphical Models and Belief Propagation, as well as to researchers in quantum theory, particularly in quantum information and the simulation of quantum many-body systems.  As such, it is intended to be as self-contained as possible, although we do assume familiarity with the basic formalism of quantum theory on finite dimensional Hilbert spaces, including the theory of density matrices, generalized measurements and completely positive maps, as used in quantum information theory.  These are covered in detail in the textbook of Nielsen and Chuang~\cite{NC00a}, as well as in Preskill's lecture notes~\cite{Pre99b}.  For further background on classical Graphical Models and Belief Propagation, we suggest the texts of Lauritzen~\cite{Lau96a}, MacKay~\cite{Mac03a}, and Neapolitan~\cite{Nea90a,Nea04a}, as well as the review articles by Yedida et al.~\cite{Yed01a,YFW02a} and Aji and McEliece~\cite{AM00a}.

The remainder of this paper is structured as follows.  In \S\ref{Problem}, the generic classical and quantum probabilistic inference problems are defined.  In \S\ref{Cond}, we review the notions of classical and quantum conditional independence, which are crucial for the development of Graphical Models and Belief Propagation algorithms.  \S\ref{Cond:CMI} outlines the entropic approach to conditional independence based on the vanishing of conditional mutual information and the associated constraints on conditional and mutual probability distributions.  This entropic approach has a straightforward quantum generalization based on the equality conditions for strong-subadditivity, which is described in \S\ref{Cond:QCMI}.  \S\ref{Cond:CDO} introduces the quantum conditional and mutual density operators, which are analogous to classical conditional and mutual probability distributions, and \S\ref{Cond:QCI} explains how quantum conditional independence can be characterized directly in terms of them.    

In \S\ref{Graph}, we develop the theory of quantum Graphical Models.    \S\ref{Graph:MN} reviews the definition of classical Markov Networks and the Hammersley-Clifford theorem, which gives an explicit representation of the probability distributions supported on them.  Motivated by this, \S\ref{Graph:QGS} defines the class of quantum $n$-Bifactor Networks, which are the most general class of networks on which our Belief Propagation algorithms operate. \S\ref{Graph:DMG} reviews the theory of dependency models and graphoids, which are abstractions of the conditional independence relation, and a quantum graphoid is defined based on quantum conditional independence. \S\ref{Graph:HC} uses the quantum graphoid to define quantum Markov Networks and gives some partial characterization theorems, along the lines of the classical Hammersley-Clifford theorem, which connect quantum Markov Networks to $n$-Bifactor Networks.  \S\ref{Graph:OM} briefly discusses quantum generalizations of two other classical Graphical Models: Factor Graphs and Bayesian Networks.  Figure \ref{fig:worldview} sketches the relation between some of these Graphical Models, and summarizes the Quantum Belief Propagation algorithm's domain of convergence.  

\S\ref{QBP} discusses the Quantum Belief Propagation algorithms.  In \S\ref{QBP:Desc}, QBP algorithms are described for $n$-Bifactor Networks.  In \S\ref{QBP:Conv}, QPB is shown to converge for $1$-Bifactor Networks on trees and for general Bifactor Networks on trees that are also Quantum Markov Networks.  \S\ref{Heur}  discusses some methods for using QBP as a heuristic algorithm in cases where it is not known to converge.  These are coarse graining \S\ref{Heur:CG}, sliding window QBP \S\ref{Heur:SW} and the method of replicas \S\ref{Heur:Rep}.  

\S\ref{App} presents two applications of QBP: to decoding quantum error correcting codes in \S\ref{App:QEC} and to simulating many-body quantum systems in \S\ref{App:Stat}.  In particular,  \S\ref{App:Stat} explains how projected entangled-pair states, which have been successfully used in statistical physics as approximations to the ground states of a wide class of Hamiltonians, can be incorporated into the framework of Bifactor Networks. 

To conclude, \S\ref{Relate} discusses the relationship to other quantum generalizations of Graphical Models and Belief Propagation that have been proposed and \S\ref{Conc} describes open questions and future research directions suggested by this work.

Note that a slightly unconventional notation for probability distributions on sets of random variables and for quantum states on tensor products of quantum systems is used throughout.  This is very convenient for describing Graphical Models and is reviewed in appendix \ref{ProbNot}.  

\section{Classical and Quantum Probabilistic Inference}

\label{Problem}

Classical Graphical Models are designed to be used as tools for making probabilistic inferences amongst large numbers of correlated random variables.  Consider a set random variables, $V = \{v_1, v_2, \ldots, v_N\}$, each of which takes a finite number of integer values $\{1,2,\ldots d\}$.  To specify a general probability distribution, $P(V)$, over the variables requires $O(d^N)$ parameters.  On learning that some subset of the variables $U \subseteq V$ take particular values, denoted $\tilde{U} = \{u = j_u\}_{u \in U}$, an important task is to update the probability for some other disjoint subset of variables $W \subseteq V$ via Bayes rule
\begin{equation}
\label{Intro:Bayes}
P(W|\tilde{U}) = \frac{P(\tilde{U}\cup W)}{P(\tilde{U})} = \frac{\sum_{V - (U \cup W)}P(\tilde{U} \cup (V - U))}{\sum_{V - U} P(\tilde{U}\cup (V - \tilde{U}))}.
\end{equation}
This immediately raises two problems.  Firstly, the number of parameters needed to specify the input to the computation, i.e. the probability distribution itself, is exponential in $N$.  We would like to specify a well-defined computational problem in which $N$ measures the input size.  Therefore, it is not feasible to consider the full set of probability distributions over $N$ variables, and attention must be restricted to families of distributions that can be specified with a number of parameters that grows only polynomially in $N$.  Secondly, assuming that the sizes of $U$ and $W$ are held constant as $N$ increases, eq.~\eqref{Intro:Bayes} involves sums over a number of terms that is exponential in $N$.  Thus, a straightforward evaluation of the formula would not give an efficient algorithm.  The restriction on the class of probability distributions must somehow be used to find an alternative method of computation that is efficient.

Classical Graphical Models are designed to provide an efficient representation of classes of probability distributions and Belief Propagation algorithms are designed to solve the corresponding inference problem.

In quantum theory, the random variables are replaced by a set of $N$ quantum systems $V = \{v_1,v_2,\ldots,v_N\}$, each associated with a Hilbert space of dimension $d$.   Again, it takes an exponential in $N$ number of parameters to specify a general density operator $\rho_V$.  The analog of the inference in eq.~\eqref{Intro:Bayes} is to perform a positive operator valued measure (POVM) $\left \{ E_U^{(j)}\right \}$ on a subsystem $U \subseteq V$ and, on obtaining outcome $j$, update the state of some disjoint subsystem $W \subseteq V$ according to
\begin{equation}
\label{Intro:QBayes}
\rho_{U| E_U^{(j)}} = \frac{\PTr{U}{E_U^{(j)}  \rho_{U \cup W}}}{\Tr{E_U^{(j)}\rho_{U}}} = \frac{\PTr{V - W}{E_U^{(j)}  \rho_{V}}}{\Tr{E_U^{(j)} \rho_{V}}}.
\end{equation}

It should be noted that this quantum problem reduces to the classical case when all the operators involved commute and are diagonal in a product basis of the systems in $V$.  In this sense eq.~\eqref{Intro:QBayes} is a noncommutative generalization of eq.~\eqref{Intro:Bayes} and this correspondence provides the guiding principle that we use to generalize the classical theory. 

The quantum problem raises the same sort of issues as in the classical case, since it takes an exponential in $N$ number of parameters to specify a state on $N$ subsystems and the trace and partial trace in eq.~\eqref{Intro:QBayes} involve sums over an exponential number of terms.  In quantum many-body theory, physical considerations are often used to motivate solutions to the representation problem, e.g. we may restrict attention to the ground or Gibbs states of some class of efficiently specifiable Hamiltonians.  In this paper, we take a different approach and instead generalize the sort of constraints that are used in defining classical Graphical Models.  The reasons for this are twofold.  Firstly, with the advent of quantum information science, it is relevant to solve instances of eq.~\eqref{Intro:QBayes} that are of broader scope than those typically considered in statistical physics.  For example, we may be interested in states that are the output of a class of polynomial quantum circuits, or in the code states of a quantum error correction code.  The most natural way to phrase such constraints is not always in terms of Hamiltonians, although it may be possible to do so.  Secondly, by focussing on constraints with a clear probabilistic and information theoretic meaning, the connection between the classical and quantum problems is elucidated and the results of the vast literature on the classical inference problem can be called into play.

% ***** 2) CONDITIONAL INDEPENDENCE *****

\section{Conditional Independence}

\label{Cond}

The formal construction of classical Graphical Models is based on the idea of placing conditional independence constraints on sets of random variables.  In this section, the relevant classical definitions are reviewed and their quantum generalizations are introduced.  In \S\ref{Cond:CMI}, the entropic approach to conditional independence is outlined and the corresponding constraints on conditional and mutual probability distributions are reviewed.  In \S\ref{Cond:QCMI}, the entropic definition is straightforwardly generalized to the quantum case by replacing the Shannon entropy with the von Neumann entropy.  In order to provide constraints on density operators that are analogous to those for classical conditional and mutual probability distributions, conditional and mutual density operators are defined in \S\ref{Cond:CDO} and quantum conditional independence is expressed in terms of them in \S\ref{Cond:QCI}. 

\subsection{Classical Conditional Independence}

\label{Cond:CMI}

For a set $V$ of classical random variables with joint distribution $P(V)$, the marginal distribution for any $U \subseteq V$ is defined as $P(U) = \sum_{V-U} P(V)$
and for any two disjoint sets $U,W \subseteq V$, the conditional distribution of $U$ given $W$ is defined as
\begin{equation}
P(U|W) = \frac{P(U \cup W)}{P(W)}
\end{equation}
The Shannon entropy of any $U \subseteq V$ is defined as
\begin{equation}
H(U) = -\sum_{U} P(U) \log_2 P(U)
\end{equation}
For disjoint $U,W \subseteq V$, the conditional entropy of $U$ given $W$ is defined as
\begin{equation}
\label{Cond:SECond}
H(U|W) = - \sum_{U \cup W} P(U \cup W) \log_2 P(U|W),
\end{equation}
and satisfies the identity
\begin{equation}
\label{Cond:SEChain}
H(U|W) = H(U \cup W) -  H(W).
\end{equation}
The mutual information between $U$ and $W$ is defined to be
\begin{align}
H(U:W) & = H(U) - H(U|W) \label{Cond:SMut} \\
& = H(U) + H(W) - H(U \cup W). \label{Cond:SMutJ}
\end{align}
Note that $H(U:W) = 0$ iff $P(U \cup W) = P(U)P(W)$.
For three disjoint sets $U,W,X \subseteq V$, the conditional mutual information between $U$ and $W$, given $X$ is defined to be
\begin{align}
H(U:W|X) & = H(U|X) - H(U|W \cup X) \\
& = H(U \cup X) + H(W \cup X) - H(X)  - H(U \cup W \cup X). \label{Cond:SCMutJ}
\end{align}
The condition $H(U:W|X) = 0$ is known as \emph{conditional independence} of $U$ and $W$ given $X$ and it is equivalent to any of the following conditions
\begin{align}
P(U|W \cup X) & = P(U|X) \label{Cond:CI1} \\
P(W|U \cup X) & = P(W|X) \label{Cond:CI2}\\
P(U\cup W|X) & = P(U|X)P(W|X) \label{Cond:CI3} \\
P(U \cup W \cup X) & = P(U|X)P(W|X)P(X). \label{Cond:CI4}
\end{align}

\begin{Exa}
\label{Cond:MC}
Consider a Markov chain consisting of three random variables $u - x - w$.  The defining condition for such a process is that $u$ and $w$ are conditionally independent given $x$.  Thus, eq. \eqref{Cond:CI4} immediately implies that the joint probability distribution has the form 
\begin{equation}
\label{Cond:CCause}
P(u,x,w) = P(u|x)P(w|x)P(x).
\end{equation}
In general, a joint distribution of three variables can be written as $P(u,x,w) = P(w|u,x)P(x|u)P(u) = P(u|x,w)P(x|w)P(w)$ and so eqs. (\ref{Cond:CI1}) and (\ref{Cond:CI2}) imply that $P(u,x,w)$ can also be written as
\begin{align}
P(u,x,w) & = P(w|x)P(x|u)P(u) \label{Cond:Forward} \\
P(u,x,w) & = P(u|x)P(x|w)P(w). \label{Cond:Backwards}
\end{align}
The three equivalent decompositions given in eqs. (\ref{Cond:CCause} - \ref{Cond:Backwards}) are suggestive of three different types of causal scenario that might give rise to such a Markov chain: 
\begin{align*}
(\ref{Cond:CCause}) \,\, \text{suggests} \,\, x \,\, & \text{is a common cause of} \,\, u \,\, \text{and}\,\,  w:   \qquad u \leftarrow x \rightarrow w \\
(\ref{Cond:Forward}) \,\, \text{suggests} \,\, u \,\, & \text{causes}\,\, x \,\, \text{and then}\,\, x \,\, \text{causes}\,\, w:   \qquad u \rightarrow x \rightarrow w \\  
(\ref{Cond:Backwards}) \,\, \text{suggests} \,\, w \,\, & \text{causes}\,\, x \,\,\text{and then}\,\, x \,\, \text{causes}\,\, u:    \qquad u \leftarrow x \leftarrow w.
\end{align*}
The common feature of these three scenarios is that in each case all the correlations between $u$ and $w$ are mediated by $x$.  Ultimately, conditional independence captures this common feature rather than implying any specific causal scenario.
\end{Exa}

The example shows that care should be taken when interpreting a decomposition of a joint probability distribution into conditional and marginal distributions.  Conditional independence is about the structure of correlations between random variables rather than their specific causal relations.  For this reason it is often useful to replace conditional probabilities with an object that is more closely connected with correlation. 

The \emph{mutual probability distribution} of disjoint $U, W \subseteq V$ is given by
\begin{equation}
\label{Cond:MPD}
P(U:W) = \frac{P(U\cup W)}{P(U)P(W)} = \frac{P(U|W)}{P(U)}.
\end{equation}
As the name implies, this is related to the mutual information and it is easy to check that eq. \eqref{Cond:SMut} can be rewritten as
\begin{equation}
\label{Cond:SMIMutForm}
H(U:V) = \sum_{U \cup W} P(U,W) \log_2 P(U:W).
\end{equation}

The conditional independence conditions eqs. (\ref{Cond:CI1}-\ref{Cond:CI4}) can be re-expressed in terms of mutual distributions as
\begin{align}
P(U:W \cup X) & = P(U:X) \label{Cond:MI1} \\
P(W:U \cup X) & = P(W:X) \label{Cond:MI2}\\
P(U\cup W:X) & = P(U:X)P(W:X) \label{Cond:MI3} \\
P(U \cup W \cup X) & = P(U:X)P(W:X)P(X)P(U)P(W). \label{Cond:MI4}
\end{align}

\begin{Exa}
Returning to the Markov chain of example \ref{Cond:MC}, the decompositions eqs. (\ref{Cond:CCause}-\ref{Cond:Backwards}) can all be rewritten in terms of mutual distributions by replacing each conditional probability with the product of a marginal and a mutual distribution using the relation $P(U|W) = P(U:W)P(U)$.  All three decompositions reduce to the same expression:
\begin{equation}
\label{Cond:MCMDecomp}
P(u,x,w) = P(u)P(x)P(w)P(u:x)P(x:w).
\end{equation}
This decomposition clearly shows that all correlations between $u$ and $w$ are mediated by $x$ and avoids the causal ambiguities that are implicit in the use of conditional probabilities.
\end{Exa}

\subsection{Quantum Conditional Independence}

\label{Cond:QCMI}

Turning now to the quantum case, if $V$ is a set of subsystems then the joint state is a density operator $\rho_V$.  For $U \subseteq V$, the analog of a marginal distribution is the reduced state obtained by taking a partial trace over $V - U$, i.e. $\rho_U = \PTr{V-U}{\rho_V}$.  The Shannon entropy is replaced by the von Neumann entropy, defined as
\begin{equation}
S(U) = -\Tr{\rho_U \log_2 \rho_U}.
\end{equation}
Quantum analogs of conditional and mutual probability distributions are not commonly discussed in the literature, but they are needed to obtain decompositions of the joint density operator analogous to eqs. (\ref{Cond:CI1}-\ref{Cond:CI4}) and eqs. (\ref{Cond:MI1}-\ref{Cond:MI4}), so they are introduced in the next section.  For now, note that the quantum conditional entropy, mutual information and conditional mutual information can already be defined by simply replacing $H$ with $S$ in the expressions \eqref{Cond:SEChain}, \eqref{Cond:SMutJ} and \eqref{Cond:SCMutJ}, since these expressions only involve joint and marginal probability distributions.  

By comparison with the classical case, it is natural to consider $S(U:W|X) = 0$ as a definition of \emph{quantum conditional independence}.  In fact, the inequality $S(U:W|X) \geq 0$ always holds and is known as strong subadditivity, so quantum conditional independence is simply the equality condition for strong subadditivity.  This equality condition has been investigated extensively and has been shown \cite{HJPW03a} to be equivalent to the existence a decomposition of the Hilbert space $\mathcal{H}_X$ of the form
\begin{equation}
\mathcal{H}_X  = \bigoplus_{j = 1}^d \left ( \mathcal{H}_{X_j^L} \otimes \mathcal{H}_{X_j^R} \right ),
\label{eq:decomp_X}
\end{equation}
(the superscripts $L$ and $R$ indicate the left and right sector of the tensor product) such that the joint density operator $\rho_{U\cup W \cup X}$ can be written as
\begin{equation}
\label{Cond:Hayden}
\rho_{U \cup W \cup X} = \sum^d_{j = 1} p_j \sigma_{U X_j^L} \otimes \tau_{X_j^R W},
\end{equation}
where $0 \leq p_j \leq 1$, $\sum_{j = 1}^d p_j = 1$, and $ \sigma_{U X_j^L}$ and $\tau_{X_j^R W}$ are density operators on $\mathcal{H}_{U} \otimes \mathcal{H}_{X_j^L}$ and $\mathcal{H}_{X_j^R} \otimes \mathcal{H}_{W}$ respectively.

Less explicit formulations of the equality condition have also been found \cite{Rus02b}, such as the operator equality
\begin{equation}
\label{Cond:Ruskai}
\log \rho_{UWX} + \log \rho_X = \log \rho_{UX} + \log \rho_{WX},
\end{equation}
where the logarithms are restricted to the supports of the operators.

\subsection{Conditional and Mutual Density Operators}

\label{Cond:CDO}

Quantum conditional independence can be expressed in a form closer to the classical conditions eqs.~(\ref{Cond:CI1}-\ref{Cond:CI4}) and (\ref{Cond:MI1}-\ref{Cond:MI4}) by introducing definitions of \emph{conditional and mutual density operators}.  For this purpose, it is convenient to define a family of products for pairs of operators $A, B$ as follows. 
\begin{equation}
A \pns{n} B = \left ( A^{\frac{1}{2n}} B^{\frac{1}{n}} A^{\frac{1}{2n}} \right )^n
\end{equation}
An important property of the $\pns{n}$ products is that if $A$ and $B$ are both positive operators then $A \pns{n} B$ is also positive.  In what follows, the most frequently used of these products are $A \po B = A \pns{1} B$ and
\begin{equation}
A \pity B = \lim_{n \rightarrow \infty} \left ( A \pns{n} B \right ).
\end{equation}
Note that whilst $\pity$ is commutative and associative, $\pns{n}$ is neither in general, so particular attention must be paid to the ordering of operators.  

The product $\pity$ was previously introduced in \cite{War05a}, in the context of a Bayesian calculus for quantum theory, and it satisfies the formula
\begin{equation}
\label{Cond:LogDef}
A \pity B = \exp \left ( \log A + \log B \right ),
\end{equation}
whenever $A$ and $B$ are strictly positive.  If $A$ and $B$ are semi-positive, then eq.~\eqref{Cond:LogDef} may be extended by restricting the action of the logarithm to the supports of the operators.

The $\pns{n}$ products can be used to define a family of conditional density operators.  Let $V$ be a set of quantum systems in a state $\rho_V$ and let $U,W \subseteq V$ be disjoint. Define
\begin{equation}
\rho\ns{n}_{U|W} = \rho_W^{-1} \pns{n} \rho_{U\cup W}, 
\end{equation}
where $^{-1}$ denotes the Moore-Penrose pseudoinverse\footnote{In the present case this means that $\rho_W^{-1}$ is the inverse of $\rho_W$ when restricted to the support of $\rho_W$ and has the same null space as $\rho_W$.}.  Note that if $W = \emptyset$, so that $\mathcal{H}_W =\mathbb{C}$ is the trivial Hilbert space, then $\rho\ns{n}_{U|W} = \rho_U$.  The conditional density operators used most frequently in this paper are $\rho\1_{U|W} = \rho\ns{1}_{U|W}$ and $\rho\ity_{U|W} = \rho_W^{-1}  \pity \rho_{U\cup W}$.

The operator $\rho\ity_{U|W}$ was originally introduced \cite{CA97a} because it allows the quantum conditional entropy to be expressed via a formula analogous to eq.~\eqref{Cond:SECond}
\begin{equation}
S(U|W) = -\Tr{\rho_{U\cup W} \log_2 \rho\ity_{U|W}}.
\end{equation}
The operator $\rho\1_{U|W}$ was introduced in \cite{Lei06a, Lei06b, AKMS06a} and also exhibits strong analogies with classical conditional probability. 

The corresponding family of \emph{mutual density operators} is defined similarly via
\begin{equation}
\label{Cond:MDODef}
\rho^{(n)}_{U:W} = \left ( \rho_U^{-1} \otimes \rho_W^{-1} \right) \po^{(n)} \rho_{U\cup W} = \rho_U^{-1} \po^{(n)} \rho^{(n)}_{U|W},
\end{equation}
with $\rho\ity_{U:W}$ and $\rho\1_{U:W}$ defined in the obvious way.  

The operator $\rho\ity_{U:W}$ was introduced \cite{CA97a} in order to express the quantum mutual information via a formula analogous to eq.~\eqref{Cond:SMIMutForm}
\begin{equation}
S(U:W) = -\Tr{\rho_{U \cup W} \log_2 \rho\ity_{U:W}}.
\end{equation}

\subsection{Constraints on Conditional and Mutual Density Operators}

\label{Cond:QCI}

In this section, quantum conditional independence is shown to be equivalent to constraints on conditional and mutual density operators analogous to eqs.~(\ref{Cond:CI1}-\ref{Cond:CI4}) and eqs.~(\ref{Cond:MI1}-\ref{Cond:MI4}). 

\begin{The}
If $S(U:W|X) = 0$ then the following conditions hold:
\begin{align}
\rho\ns{n}_{U|X\cup W} & = \rho\ns{n}_{U|X} \otimes P_W \label{Cond:CIo1}\\
\rho\ns{n}_{W|X \cup U} & = \rho\ns{n}_{W|X} \otimes P_U \label{Cond:CIo2}\\
\rho\ns{n}_{U \cup W |X} & = \rho\ns{n}_{U|X} \rho\ns{n}_{W|X}\label{Cond:CIo3}\\
\rho_{U\cup W \cup X} & = \rho_X \pns{n} \left ( \rho\ns{n}_{U|X} \rho\ns{n}_{W|X} \right ), \label{Cond:CIo4} 
\end{align}
where $P_W$ is the projector onto the support of $\rho_W$ and $P_U$ is the projector onto the support of $\rho_U$.
\end{The}
\begin{proof}
These conditions are a direct consequence of the decomposition given in eq. \eqref{Cond:Hayden}.  Since each $\mathcal{H}_{X_j^L}$ is a factor in a direct sum decomposition of $\mathcal{H}_X$, it follows that the operators $\sigma_{UX_j^L}$ have disjoint support.  Similarly, the operators $\tau_{W X_j^R}$ have disjoint support.  Hence, to prove eq. \eqref{Cond:CIo1} note that
\begin{equation}
\rho_{W \cup X} = \sum^d_{j = 1} p_j \sigma_{X_j^L} \otimes \tau_{X_j^R W},
\end{equation} 
and hence
\begin{align}
\rho\ns{n}_{U|W \cup X}  & = \rho_{WX}^{-1} \pns{n} \rho_{UWX} \\
& = \sum^d_{j = 1} \left ( \sigma_{X_j^L}^{-1} \pns{n} \sigma_{U X_j^L} \right ) \otimes \left ( \tau_{X_j^R W}^{-1} \pns{n} \tau_{X_j^R W} \right ) \\
& = \sum^d_{j = 1} \sigma\ns{n}_{U |X_j^L} \otimes P_{X_j^R W} \\
& = \rho\ns{n}_{U|X} \otimes P_W,
\end{align}
where $P_{X_j^R W}$ is the projector onto the support of $\tau_{X_j^R W}$.

Eqs. \eqref{Cond:CIo2} and \eqref{Cond:CIo3} are proved similarly, with the proviso that the decomposition given in eq. \eqref{Cond:Hayden} implies that $\rho\ns{n}_{U|X}$ and $\rho\ns{n}_{W|X}$ commute, which is necessary to prove eq. \eqref{Cond:CIo3}.  Finally, \eqref{Cond:CIo4} is equivalent to \eqref{Cond:CIo3} via the definition a conditional density operator.
\end{proof}

It is straightforward to adapt the proof in order to arrive at analogous decompositions in terms of mutual density operators.
\begin{The}
If $S(U:W|X) = 0$ then the following conditions hold:
\begin{align}
\rho\ns{n}_{U:X\cup W} & = \rho\ns{n}_{U:X} \otimes P_W \label{Cond:MIo1}\\
\rho\ns{n}_{W:X \cup U} & = \rho\ns{n}_{W:X} \otimes P_U \label{Cond:MIo2}\\
\rho\ns{n}_{U \cup W :X} & = \rho\ns{n}_{U:X} \rho\ns{n}_{W:X}\label{Cond:MIo3}\\
\rho_{U\cup W \cup X} & = \left ( \rho_U \otimes \rho_W \otimes \rho_X \right ) \pns{n} \left ( \rho\ns{n}_{U:X} \rho\ns{n}_{W:X} \right ), \label{Cond:MIo4}
\end{align}
\end{The}

% Comments to be reconsidered in light of proof.
%The conditions eq.~\eqref{Cond:CIo3} and \eqref{Cond:CIo4} are surprising because a product of positive operators is not positive in general.  However, it follows from eq.~\eqref{Cond:Hayden} that $\rho\ns{n}_{U|X}$ and $\rho\ns{n}_{W|X}$ commute, in which case their product is positive and is in fact equal to $\rho\ns{n}_{U|X} \pns{n} \rho\ns{n}_{W|X}$. It also follows from eq.~\eqref{Cond:Hayden} that $\rho_X$ is decomposable with respect to the pair $\rho\ns{n}_{U|X}$ and $\rho\ns{n}_{W|X}$. 

It remains to determine whether any converse implications hold, i.e. which of the conditions eqs. (\ref{Cond:CIo1}-\ref{Cond:CIo4}) and (\ref{Cond:MIo1}-\ref{Cond:MIo4}) imply that $S(U:W|X) = 0$.  For this purpose, it is only necessary to consider eqs. (\ref{Cond:CIo1}-\ref{Cond:CIo3}) because eqs. (\ref{Cond:MIo1}-\ref{Cond:MIo4}) are equivalent to eqs. (\ref{Cond:CIo1}-\ref{Cond:CIo4}) via the definition of a mutual density operator and eq. \eqref{Cond:CIo4} is equivalent to eq. \eqref{Cond:CIo3} via the definition of a conditional density operator.  In general, the situation appears to be more complicated than in the classical case and we are only able to obtain tight converse results for the cases $n \rightarrow \infty$ and $n = 1$.

\begin{The}
\label{Cond:RuskaiThe}
In the limit, $n \rightarrow \infty$, all the converse implications hold, i.e. any of the conditions (\ref{Cond:CIo1}-\ref{Cond:CIo3}) imply that $S(U:W|X) = 0$. 
\end{The} 
\begin{proof}
These results are simple consequences of the equality condition given in eq. \eqref{Cond:Ruskai}.  For eq. \eqref{Cond:CIo1} we have
\begin{equation}
\rho_{W \cup X}^{-1} \pity \rho_{U \cup W \cup X} = \rho_{X}^{-1} \pity \rho_{U \cup X}.
\end{equation}
Using eq. \eqref{Cond:LogDef} gives
\begin{equation}
\exp \left ( \log \rho_{U\cup W \cup X} - \log \rho_{W \cup X }\right ) = \exp \left ( \rho_{U \cup X} - \rho_X \right ).
\end{equation}
Taking logarithms and rearranging gives eq. \eqref{Cond:Ruskai}.  The proofs for eqs. \eqref{Cond:CIo2} and \eqref{Cond:CIo3} follow by similar arguments.
\end{proof}

For the $n=1$ case, eqs. \eqref{Cond:CIo1} and \eqref{Cond:CIo2} imply converse results.  

\begin{The}
If $\rho\1_{U|X\cup W}  = \rho\1_{U|X}$ or $\rho\1_{W|X \cup U} = \rho\1_{W|X}$ then $S(U:W|X) = 0$. 
\end{The}
\begin{proof}
As explained in \cite{HJPW03a}, Uhlman's theorem \cite{Uhl77a}, implies that $S(U:W|X) = 0$ iff there exists a trace preserving, completely positive map $\mathcal{E}_{U\cup X \cup W|U \cup X}: \mathfrak{L}(\mathcal{H}_{U} \otimes \mathcal{H}_X) \rightarrow \mathfrak{L}(\mathcal{H}_{U} \otimes \mathcal{H}_X \otimes \mathcal{H}_W)$, such that both
\begin{align}
\mathcal{E}_{U \cup X \cup W|U \cup X}(\rho_U \otimes \rho_X) & =  \rho_{U} \otimes \rho_{X \cup W} \\
\mathcal{E}_{U \cup X \cup W|U \cup X}(\rho_{U \cup X}) & =  \rho_{U \cup X \cup W}
\end{align}
hold simultaneously.  In the present case, this can be achieved via a map of the form $\mathcal{E}_{U \cup X \cup W|U \cup X} = \mathcal{I}_U \otimes \mathcal{F}_{X \cup W|X}$, where $\mathcal{I}_U$ is the identity superoperator on $\mathfrak{L}(\mathcal{H}_U)$ and $\mathcal{F}_{X\cup W|X}: \mathfrak{L}(\mathcal{H}_X) \rightarrow \mathfrak{L}(\mathcal{H}_X \otimes \mathcal{H}_W)$ is a trace preserving completely positive map.  $\mathcal{F}_{X \cup W|X}$ is defined via a Kraus representation $\mathcal{F}_{X \cup W|X}(\sigma_X) = \sum_j M^{(j)}_{X \cup W|X} \sigma_X M^{(j)\dagger}_{X \cup W|X}$, where
\begin{equation}
M^{(j)}_{X \cup W|X} = \rho_{X \cup W}^{\frac{1}{2}} \Ket{j}_W \rho_{X}^{-\frac{1}{2}},
\end{equation}
and $\Ket{j}_W$ are basis vectors for $\mathcal{H}_W$.

It is straightforward to check that $\sum_j M^{(j)\dagger}_{X \cup W|X} M^{(j)}_{X \cup W|X} = P_X$, where $P_X$ is the projector onto the support of $\rho_X$.  This can easily be extended to be a trace preserving map by adding an extra Kraus operator that has support only in the subspace orthogonal to the support of $\rho_X$, but this can be omitted for the present purpose since it doesn't change the action of $\mathcal{E}_{U \cup X \cup W|U \cup X}$ on $\rho_U \otimes \rho_X$ or $\rho_{U \cup X}$.  It is straightforward to check that $\mathcal{F}_{X \cup W|X} ( \rho_X ) = \rho_{X \cup W}$, so the first condition is satisfied.  The action on $\rho_{U \cup X}$ is given by
\begin{align}
\mathcal{I}_U \otimes \mathcal{F}_{X \cup W|X} (\rho_{U \cup X}) & = \sum_j I_U \otimes M^{(j)}_{X \cup W|X} \rho_{U \cup X} I_U \otimes M^{(j)}_{X \cup W|X} \\
& = \rho_{X \cup W}^{\frac{1}{2}} \sum_j \Ket{j}_W \Bra{j}_W \rho_{X}^{-\frac{1}{2}} \rho_{U \cup X} \rho_{X}^{-\frac{1}{2}} \rho_{X \cup W}^{\frac{1}{2}}  \\
& = \rho_{X \cup W}^{\frac{1}{2}} \rho\1_{U|X} \rho_{X \cup W}^{\frac{1}{2}}
\end{align}
By assumption, $\rho\1_{U|X\cup W}  = \rho\1_{U|X}$, so it follows that $\rho_{U \cup X \cup W} = \rho_{X \cup W}^{\frac{1}{2}} \rho\1_{U|X} \rho_{X \cup W}^{\frac{1}{2}}$, as required.  The result for  $\rho\1_{W|X \cup U} = \rho\1_{W|X}$ follows by symmetry.
\end{proof}

For $n<\infty$, it is not true that \eqref{Cond:CIo3} implies conditional independence, even in the case $n=1$. This is illustrated by the following counterexample. 

\begin{Exa}
\label{ex:notMarkov}
Let $U$ and $W$ be single qubits, and $X$ be composed of  two qubits labeled $X^L$ and $X^R$. For $\epsilon >0$, consider the normalized state
\begin{equation}
\rho_{U\cup X \cup W} = \frac{4}{(1-\epsilon)^{\frac 1n} + 3(\epsilon/3)^{\frac 1n}}
\rho_X \pns{n} \Big(P^-_{U\cup X^L} \otimes P^-_{W\cup X^R}\Big),
\end{equation}
where
\begin{equation}
\rho_X = (1-\epsilon) P^-_{X^L\cup X^R} + \frac{\epsilon}{3} P^+_{X^L\cup X^R}
\end{equation}
and where $P_{A\cup B}^\pm$ denote the projector onto the symmetric and anti-symmetric subspaces of $\mathcal{H}_A \otimes \mathcal{H}_B$. The conditional states are 
\begin{align}
\rho_{U| X}^{(n)} &= \frac{2}{\sqrt{(1-\epsilon)^{\frac 1n} + 3(\epsilon/3)^{\frac 1n}}} P^-_{U\cup X^L} \otimes I_{X^R} \ \mathrm{and} \\ 
\rho_{W| X}^{(n)} &= \frac{2}{\sqrt{(1-\epsilon)^{\frac 1n} + 3(\epsilon/3)^{\frac 1n}}} I_{X^L} \otimes P^-_{W\cup X^R}.
\end{align}   
By construction, condition \eqref{Cond:CIo4}  is easily verified $\rho_{U\cup X\cup W} = \rho_X \pns{n} (\rho_{U| X}^{(n)} \rho_{W| X}^{(n)})$. In the limit $\epsilon \rightarrow 0$, the state $\rho_{U\cup X \cup W} \rightarrow P^-_{U\cup W} \otimes P^-_{X^L\cup X^R}$, which has $S(U:W|X) = 2$. By continuity, we claim that  for all $n<\infty$, there exists an $\epsilon > 0$ such that $\rho_{U\cup X\cup W}$ is a density operator that does not saturate strong subadditivity. 
\end{Exa}

The preceding example shows that some of the conditions given in eqs. (\ref{Cond:CIo1}-\ref{Cond:CIo4}) are not sufficient to imply quantum conditional independence on their own.   Therefore, additional constraints need to be imposed in order to obtain converse results.  Two alternative approaches are considered here, one based on additional commutation conditions that hold for conditionally independent states and one based on the algebraic structure of such states. The approach based on commutation conditions is perhaps more elegant, but the algebraic conditions are also relevant because they are used in theorem \ref{Graph:THCC} in \S\ref{Graph:HC} to provide a characterization result for quantum Markov Networks on trees.  The following sequence of results provides the approach based on commutation conditions.

\begin{The}
\label{Cond:EqThe}
For a fixed $n$, if $\rho_X^{-\frac{1}{2n}}\rho_{U\cup X}^{\frac{1}{2n}}$ and its adjoint commute with $\rho_X^{-\frac{1}{2n}}\rho_{W \cup X}^{\frac{1}{2n}}$, then the conditions given in eqs. (\ref{Cond:CIo1}-\ref{Cond:CIo4}) are all equivalent.
\end{The}
\begin{proof}
We start by showing that $\rho\ns{n}_{U|W\cup X} = \rho\ns{n}_{U|X}$ is equivalent to $\rho\ns{n}_{W|U \cup X} = \rho\ns{n}_{W|X}$.  The first of these can be written explicitly in terms of joint and reduced density operators as
\begin{equation}
\rho_{W \cup X}^{-\frac{1}{2n}} \rho_{U \cup W \cup X}^{\frac{1}{n}} \rho_{W \cup X}^{-\frac{1}{2n}} = \rho_{X}^{-\frac{1}{2n}} \rho_{U \cup X}^{\frac{1}{n}} \rho_{X}^{-\frac{1}{2n}}.
\end{equation}
Left and right multiplying by $\rho_{W\cup X}^{\frac{1}{2n}}$ gives
\begin{equation}
\label{Cond:WOIF1}
\rho_{U\cup W\cup X}^{\frac{1}{n}} = \rho_{W\cup X}^{\frac{1}{2n}} \rho_{X}^{-\frac{1}{2n}} \rho_{U \cup X}^{\frac{1}{n}} \rho_{X}^{-\frac{1}{2n}} \rho_{W\cup X}^{\frac{1}{2n}}.
\end{equation}
Now, define $T = \rho_{W\cup X}^{\frac{1}{2n}} \rho_X^{-\frac{1}{2n}} \rho_{U\cup X}^{\frac{1}{2n}}$ so that $\rho_{U\cup W\cup X}^{\frac{1}{n}} = TT^\dagger$.  In a similar fashion, $\rho\ns{n}_{W|U\cup X} = \rho_{W|X}$ can be shown to be equivalent to $\rho_{U\cup W\cup X}^{\frac{1}{n}} = T^\dagger T$.  

Now,
\begin{align}
T^\dagger & = \rho_{U\cup X}^{\frac{1}{2n}} \rho_X^{- \frac{1}{2n}} \rho_{W\cup X}^{\frac{1}{2n}} \\
& =  \rho_X^{\frac{1}{2n}} \rho_X^{- \frac{1}{2n}} \rho_{U\cup X}^{\frac{1}{2n}} \rho_X^{- \frac{1}{2n}} \rho_{W \cup X}^{\frac{1}{2n}} \\
& =  \rho_X^{\frac{1}{2n}} \rho_X^{- \frac{1}{2n}} \rho_{W\cup X}^{\frac{1}{2n}} \rho_X^{- \frac{1}{2n}} \rho_{U\cup X}^{\frac{1}{2n}} \label{Cond:CommUse1} \\
& = \rho_{W\cup X}^{\frac{1}{2n}} \rho_X^{- \frac{1}{2n}} \rho_{U\cup X}^{\frac{1}{2n}} \\
& = T,
\end{align}
where the assumption that $\rho_X^{- \frac{1}{2n}} \rho_{U\cup X}^{\frac{1}{2n}}$ commutes with $\rho_X^{- \frac{1}{2n}} \rho_{W\cup X}$ has been used to derive eq. \eqref{Cond:CommUse1}.  Hence, $T$ is Hermitian and the two conditions are equivalent.

For the remaining condition note that $\rho\ns{n}_{U\cup W|X} = \rho\ns{n}_{U|X} \rho\ns{n}_{W|X}$ is equivalent to
\begin{align}
\rho_{U\cup W\cup X}^{\frac{1}{n}} & = \rho_{U\cup X}^{\frac{1}{n}} \rho_X^{-\frac{1}{n}} \rho_{W\cup X}^{\frac{1}{n}} \\
& = \rho_{U\cup X}^{\frac{1}{2n}}  \rho_{U\cup X}^{\frac{1}{2n}} \rho_X^{-\frac{1}{2n}} \rho_X^{-\frac{1}{2n}} \rho_{W\cup X}^{\frac{1}{2n}} \rho_{W\cup X}^{\frac{1}{2n}} 
\end{align}
The commutativity of $\rho_{U\cup X}^{\frac{1}{2n}} \rho_X^{-\frac{1}{2n}}$ and $\rho_X^{-\frac{1}{2n}} \rho_{W\cup X}^{\frac{1}{2n}}$ then gives
\begin{align}
\rho_{U\cup W\cup X}^{\frac{1}{n}} & = \rho_{U\cup X}^{\frac{1}{2n}}  \rho_X^{-\frac{1}{2n}} \rho_{W\cup X}^{\frac{1}{2n}} \rho_{U\cup X}^{\frac{1}{2n}} \rho_X^{-\frac{1}{2n}} \rho_X^{-\frac{1}{2n}} \rho_{W\cup X}^{\frac{1}{2n}} \\
& =  \rho_X^{\frac{1}{2n}} \rho_X^{-\frac{1}{2n}} \rho_{U\cup X}^{\frac{1}{2n}}  \rho_X^{-\frac{1}{2n}} \rho_{W\cup X}^{\frac{1}{2n}} \rho_{U\cup X}^{\frac{1}{2n}} \rho_X^{-\frac{1}{2n}} \rho_X^{-\frac{1}{2n}} \rho_{W\cup X}^{\frac{1}{2n}},
\end{align}
and the commutativity of $ \rho_X^{-\frac{1}{2n}} \rho_{U\cup X}^{\frac{1}{2n}}$ and $\rho_X^{-\frac{1}{2n}} \rho_{W\cup X}^{\frac{1}{2n}}$ gives
\begin{align}
\rho_{U\cup W\cup X}^{\frac{1}{n}} & =  \rho_X^{+\frac{1}{2n}} \rho_X^{-\frac{1}{2n}} \rho_{W \cup X}^{\frac{1}{2n}}  \rho_X^{-\frac{1}{2n}} \rho_{U\cup X}^{\frac{1}{2n}} \rho_{U\cup X}^{\frac{1}{2n}} \rho_X^{-\frac{1}{2n}} \rho_X^{-\frac{1}{2n}} \rho_{W\cup X}^{\frac{1}{2n}} \\
& =  \rho_{W\cup X}^{\frac{1}{2n}}  \rho_X^{-\frac{1}{2n}} \rho_{U\cup X}^{\frac{1}{n}}  \rho_X^{-\frac{1}{n}}  \rho_{W\cup X}^{\frac{1}{2n}},
\end{align}
which is equivalent to $\rho\ns{n}_{U|W\cup X} = \rho\ns{n}_{U|X}$.
\end{proof}

Theorem \ref{Cond:EqThe} relates the conditions eqs. (\ref{Cond:CIo1}-\ref{Cond:CIo3}) for a fixed value of $n$, but the conditions for different values of $n$ can also be related via the following corollary.

\begin{Cor}
\label{Cond:InductCor}
For fixed $n$, if  $\rho_X^{-\frac{1}{2n}}\rho_{U\cup X}^{\frac{1}{2n}}$ and its adjoint commute with $\rho_X^{-\frac{1}{2n}}\rho_{W\cup X}^{\frac{1}{2n}}$, then $\rho\ns{n}_{U|W\cup X} = \rho\ns{n}_{U|X}$ implies $\rho\ns{2n}_{U\cup W|X} = \rho\ns{2n}_{U|X}\rho\ns{2n}_{W|X}$.
\end{Cor}
\begin{proof}
In the preceding proof it was shown that $\rho\ns{n}_{U|W\cup X} = \rho\ns{n}_{U|X}$ is equivalent to $\rho_{U\cup W \cup X}^{\frac{1}{n}} = TT^{\dagger}$, where $T = \rho_{W \cup X}^{\frac{1}{2n}} \rho_X^{-\frac{1}{2n}} \rho_{U\cup X}^{\frac{1}{2n}}$, and that the commutativity conditions imply that $T$ is Hermitian.  Therefore, $\rho_{U \cup W\cup X}^{\frac{1}{n}} = \left ( T^{\dagger} \right )^2$, which implies $\rho_{U\cup W\cup X}^{\frac{1}{2n}} =  T^{\dagger} = \rho_{U\cup X}^{\frac{1}{2n}} \rho_X^{-\frac{1}{2n}} \rho_{W\cup X}^{\frac{1}{2n}}$.  The latter is straightforwardly equivalent to $\rho\ns{2n}_{U\cup W|X} = \rho\ns{2n}_{U|X}\rho\ns{2n}_{W|X}$
\end{proof}

Putting these results together leads to a set necessary and sufficient condition for conditional independence.

\begin{Cor}
If $\rho_X^{-\frac{1}{2n}}\rho_{U\cup X}^{\frac{1}{2n}}$ and its adjoint commute with $\rho_X^{-\frac{1}{2n}}\rho_{W\cup X}^{\frac{1}{2n}}$ for every $n$, then any of the conditions given in eqs. (\ref{Cond:CIo1}-\ref{Cond:CIo3}) imply that $S(U:W|X) = 0$.
\end{Cor}
\begin{proof}
Under these commutativity conditions, theorem \ref{Cond:EqThe} implies that eqs. (\ref{Cond:CIo1}-\ref{Cond:CIo3}) are equivalent for any fixed $m$ and corollary \ref{Cond:InductCor} shows that $\rho\ns{2m}_{U\cup W|X} = \rho\ns{2m}_{U|X}\rho\ns{2m}_{W|X}$ can be derived from $\rho\ns{m}_{U|W \cup X} = \rho\ns{m}_{U|X}$.  By applying theorem \ref{Cond:EqThe} with $n=2m$, it follows that $\rho\ns{m}_{U|W\cup X} = \rho\ns{m}_{U|X}$ implies $\rho\ns{2m}_{U|W\cup X} = \rho\ns{2m}_{U|X}$.  By induction, this implies that $\rho\ns{2^s m}_{U|W\cup X} = \rho\ns{2^s m}_{U|X}$ for any positive integer $s$.  Taking the limit $s \rightarrow \infty$ gives $\rho\ity_{U|W\cup X} = \rho\ity_{U|X}$, which implies $S(U:W|X) = 0$ by theorem \ref{Cond:RuskaiThe}.
\end{proof}

We now turn to the algebraic approach to proving converse results.  Firstly, note that eq.~\eqref{Cond:CIo3} implies that $\rho\ns{n}_{U|X}$ and $\rho\ns{n}_{W|X}$ commute, since $\rho\ns{n}_{U \cup W|X}$ is Hermitian.  It can be shown that whenever two operators $A_{U\cup X}\otimes I_W$ and $I_U\otimes B_{W \cup X}$ commute there exists a decomposition of $\mathcal{H}_X$ as in eq.~\eqref{eq:decomp_X} such that
\begin{align}
A_{U  X} &= \sum^d_{j = 1} a_{U X_j^L} \otimes I_{X^R_j} \ \mathrm{and} \label{Cond:Decomp1} \\
B_{W X} &= \sum^d_{j = 1}  I_{X^L_j}\otimes b_{ X_j^RW}, \label{Cond:Decomp2}
\end{align}
so eq. \eqref{Cond:CIo3} implies that $\rho\ns{n}_{U|X}$ and $\rho\ns{n}_{W|X}$ have this structure, as would be expected if the joint state is conditionally independent and hence satisfies eq.~\eqref{Cond:Hayden}.  However, eq. \eqref{Cond:Hayden} implies an additional constraint that has not been used so far, namely that $\rho_X$ also respects the same tensor product structure on $\mathcal{H}_X$, i.e. $\rho_X$ is of the form
\begin{equation}
\rho_X = \sum_{j  = 1}^d p_j \sigma_{X_j^L} \otimes \tau_{X_j^R}.
\end{equation}
More generally, we will say that an operator $C_X$ is {\em decomposable with respect to} the pair of commuting operators $A_{U\cup X}$ and $B_{W \cup X}$ if it has the same algebraic structure on $\mathcal{H}_X$, i.e. if
\begin{equation}
C_{X} = \sum^d_{j = 1} c_{X_j^R} \otimes c_{X_j^R}.
\label{def:decomposable}
\end{equation}
for some factorization of $\mathcal{H}_X$, such that eqs. \eqref{Cond:Decomp1} and \eqref{Cond:Decomp2} hold.
Imposing the commutativity of $\rho\ns{n}_{U|X}$ and $\rho\ns{n}_{W|X}$, along with the decomposability of $\rho_X$ with respect to  $\rho\ns{n}_{U|X}$ and $\rho\ns{n}_{W|X}$ as additional constraints is enough to straightforwardly show that any of eqs.~(\ref{Cond:CIo1}-\ref{Cond:CIo4}) imply conditional independence for all values of $n$. 

\section{Graphical Models}

\label{Graph}

In this section, quantum conditional independence is used to define quantum Graphical Models that generalize their classical counterparts.  The main focus is on quantum Markov Networks and $n$-Bifactor Networks, since these allow for the simplest formulation of the Belief Propagation algorithms to be described in \S\ref{QBP}.  \S\ref{Graph:MN} reviews the definition of classical Markov Networks and the Hammersley-Clifford theorem, which gives an explicit representation for the probability distributions associated with classical Markov Networks.  Motivated by this, \S\ref{Graph:QGS} defines the class of quantum $n$-Bifactor Networks, which are the most general class of networks on which our Belief Propagation algorithms operate.  \S\ref{Graph:DMG} reviews the theory of dependency models and graphoids, which is useful for proving theorems about Graphical Models, and shows that quantum conditional independence can be used to define a graphoid. \S\ref{Graph:HC} defines quantum Markov Networks and gives some partial characterization results for the associated quantum states, along similar lines to the Hammersley-Clifford theorem.  Most of these definitions and characterization results are summarized on Fig. \ref{fig:worldview}.

The remaining two subsections briefly outline two other quantum Graphical Models: Quantum Factor Graphs in \S\ref{Graph:FG} and Quantum Bayesian Networks in \S\ref{Graph:BN}.  These structures are equivalent from the point of view of the efficiency of Belief Propagation algorithms, since it is always possible to convert them into $n$-Bifactor Networks and vice-versa with only a linear overhead in graph size. An explicit method for converting a quantum factor graph into a quantum $1$-Bifactor Network is given because factor graphs are used in the application to quantum error correction developed in \S\ref{App:QEC}.

\subsection{Classical Markov Networks}

\label{Graph:MN}

\begin{figure}
\center\includegraphics{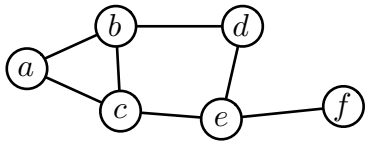}
\caption{The equalities $H(a:d\cup e\cup f | b \cup c) = 0$, $H(f:a\cup b \cup c \cup d | e) = 0$, and $H(a\cup b : e \cup f | c \cup d) = 0$ are examples of constraints that are satisfied when $(G,P(V))$ is a Markov Network.}
\label{fig:MN}
\end{figure}

Let $G = (V,E)$ be an undirected graph and suppose that each vertex $v \in V$ is associated with a random variable, also denoted $v$.  Let $P(V)$ be the joint distribution of the variables.  $(G,P(V))$ is a \emph{Classical Markov Network} if for all $U \subseteq V$, $H(U:V - (n(U) \cup U)|n(U)) = 0$, where $n(U)$ is the set of nearest neighbors of $U$ in $G$ (see Fig.~\ref{fig:MN}).  Further, if $P(V)$ is strictly positive for all possible valuations of the variables, then $(G,P(V))$ is called a \emph{Positive Classical Markov Network}.  For such positive networks there is a powerful characterization theorem \cite{Gri73a,Bes74a}.
\begin{The}[Hammersley-Clifford \cite{HC71a}]
\label{thm:HC}
$(G,P(V))$ is a positive classical Markov network iff it can be written as
\begin{equation}
\label{Graph:HCE}
P(V) = \frac{1}{Z} \prod_{C \in \mathfrak{C}} \psi(C),
\end{equation}
where $\mathfrak{C}$ is the set of cliques of $G$, $\psi(C)$ is a positive function defined on the random variables in $C$ and $Z$ is a normalization factor.
\end{The}
A set of vertices $C \subseteq V$ in a graph is a clique if $\forall u,v \in C$, $u \neq v \rightarrow (u,v) \in E$, i.e. every vertex in $C$ is connected to every other vertex in $C$ by an edge.  Note that the decomposition in eq.~\eqref{Graph:HCE} is generally not unique, even up to normalization.  A distribution of the form of eq.~\eqref{Graph:HCE} is said to factorize with respect to the graph $G$.

Markov chains are a special case of Markov Networks in which the graph is a chain.  These are included in the slightly more general class of networks where the graph is a tree.  For trees the only cliques are the individual vertices and the pairs of vertices that are connected by an edge, and the associated probability distributions have a representation in terms of marginal and mutual probability distributions of the form
\begin{equation}
\label{Graph:HCMut}
P(V) = \prod_{v \in V} P(v) \prod_{(u,v) \in E} P(u:v),
\end{equation}
which generalizes the decomposition for three variable Markov chain given in eq. \eqref{Cond:MCMDecomp}.  For more general networks wherein the graph has cycles, there is no Hammersley-Clifford decomposition in which the functions $\psi(C)$ are marginal and mutual probability distributions.

The Hammersley-Clifford decomposition can be put in a form more familiar to physicists by introducing a positive constant $\beta$ and defining the functions $H(C) = - \beta^{-1} \log \psi(C)$, which are always well defined since $\psi(C)$ is positive.  Then eq.~\eqref{Graph:HCE} can be written as
\begin{equation}
P(V) = \frac{1}{Z} \exp \left ( -\beta \sum_{C \in \mathfrak{C}} H(C) \right ),
\end{equation}
which is a Gibbs state for a system with a Hamiltonian $\sum_{C \in \mathfrak{C}} H(C)$ and partition function $Z$.  This is a generalization of the lattice models studied in statistical physics  to arbitrary graphs.  Indeed, if $G$ is a lattice, then, as for trees, the only cliques are the individual vertices and pairs of vertices connected by an edge, so for lattices the edges represent local nearest-neighbor interactions.  

In many applications,  such as in statistical physics, the functions $\psi(C)$ are often constants for cliques containing three or more vertices even in the case where the graph has cliques with more than two vertices.  In this case, we again have that the only nontrivial functions are defined on the vertices and edges of the graph, so the state can be written as
\begin{equation}
\label{Graph:CGS}
P(V) = \frac{1}{Z} \prod_{v \in V} \psi(v) \prod_{(u,v) \in E} \psi(u:v).
\end{equation}
Here, the edge functions are denoted $\psi(u:v)$ because of the close parallel with eq.~\eqref{Graph:HCMut}, but they are general positive functions rather than mutual distributions.  We adopt the terminology \emph{bifactor distribution} to describe distributions of the form of eq.~\eqref{Graph:CGS} and \emph{Bifactor Network} for the pair $(G,P(V))$.  For example, the distribution associated with a local nearest-neighbor model on an arbitrary graph, such as the spin-glasses studied in statistical physics, would be a bifactor distribution.  

\subsection{Quantum Bifactor Networks}

\label{Graph:QGS}

A proper generalization of Markov Networks to quantum theory involves the replacement of random variables with quantum systems and the replacement of classical conditional independence with its quantum counterpart.  This theory is developed in the following sections, but it is convenient to first introduce a class of states that parallels the classical bifactor distributions of eq.~\eqref{Graph:CGS}.

Let $G = (V,E)$ be a graph, let each vertex $v \in V$ be associated to a quantum system with Hilbert space $\mathcal{H}_v$.  Let $\mathcal{H}_V = \bigotimes_{v \in V} \mathcal{H}_v$ and consider the class of states $\rho_V$ that can be expressed as
\begin{equation}
\label{Graph:QGSEPre}
\rho_V = \frac 1 Z \left(\bigotimes_{u\in V} \mu_u\right) \pns{n} \left(\left (\pns{n} \right )_{(v,w) \in E} \nu_{v:w}\right),
\end{equation}
where $Z$ is normalization constant,  the $\mu_u$'s are operators on $\mathcal{H}_u$ and the $\nu_{v:w} = \nu_{w:v}$ are operators on $\mathcal{H}_v\otimes\mathcal{H}_w$.  As stated, this expression is ambiguous because the $\pns{n}$ product is neither commutative or associative apart from in the limit $n \rightarrow \infty$.  To avoid this ambiguity we impose the additional constraint that $[\nu_{u:v},\nu_{w:x}]=0$ for finite $n$, in which case the expression $\left ( \pns{n} \right )_{(v,w) \in E} \nu_{v:w}$ reduces to $\prod_{(v,w) \in E} \nu_{v:w}$. The state $\rho_V$ is an \emph{$n$-bifactor state} if it can be written as 
\begin{equation}
\label{Graph:QGSE}
\rho_V = \frac 1 Z \left(\bigotimes_{u\in V} \mu_u\right) \pns{n} \left(\prod_{(v,w) \in E} \nu_{v:w}\right),
\end{equation}
with $[\nu_{u:v},\nu_{w:x}]=0$, and it is an \emph{$\infty$-bifactor state} if it can be written as
\begin{equation}
\label{Graph:QGSEInf}
\rho_V = \frac 1 Z \left(\bigotimes_{u\in V} \mu_u\right) \pity \left(\pity_{(v,w) \in E} \nu_{v:w}\right),
\end{equation}
with no commutativity constraint on the $\nu_{v:w}$.
The pair $(G,\rho_V)$ is referred to as a quantum \emph{$n$-Bifactor Network}, or \emph{$\infty$-Bifactor Network}, respectively.

It turns out that not every quantum Bifactor Network is a quantum Markov Network, but the quantum generalizations of Belief Propagation algorithms to be developed in \S\ref{QBP} can be formulated for any Bifactor Network.  Therefore, readers who are mainly interested in algorithms and applications rather than proofs can skip to \S\ref{QBP}, perhaps pausing to read \S\ref{Graph:FG} on the way in order to understand the application to quantum error correction.  

The next goal is to formulate the theory of quantum Markov Networks and provide characterization theorems analogous to the Hammersley-Clifford theorem.  In order to do so it is convenient to first introduce the theory of dependency models and graphoids, which is useful for proving theorems about Graphical Models.

\subsection{Dependency Models and Graphoids}

\label{Graph:DMG}

Graphs and conditional independence relations share a number of important properties that are responsible for the structure of Graphical Models.  These properties are also shared by a number of other mathematical structures and they can be abstracted into structures known as dependency models and graphoids, which were introduced by Gieger, Verma, and Pearl \cite{VP90a, GVP90a}.  Here, the theory is briefly reviewed and quantum conditional independence is shown to also give rise to a graphoid.

A \emph{dependency model} $M$ over a finite set $V$ is a tripartite relation over disjoint subsets of $V$.  The statement that $(U,W,X) \in M$ will be denoted $I(U,W|X)$, with a possible subscript on the $I$ to denote the type of dependency model.  $I(U,W|X)$ should be taken to mean that ``$U$ and $W$ only interact via $X$'', or that ``$U$ and $W$ are independent given $X$''.   

\begin{Exa}
An \emph{Undirected Graph Dependency Model} $I_G$ is defined in terms of an undirected graph $G$.  Let $V$ be the set of vertices of $G$ and then let $I_G(U,W|X)$ if every path from a vertex in $U$ to a vertex in $W$ passes through a vertex in $X$.  $I_G$ is often called the \emph{Global Markov Property}.
\end{Exa}

\begin{Exa}
A \emph{Probabilistic Dependency Model} $I_P$ is defined in terms of a probability distribution $P(V)$ over a set $V$ of random variables. $I_P(U,W|X)$ is true if $U$ and $W$ are conditionally independent given $X$.  
\end{Exa}

\begin{Exa}
A \emph{Quantum Dependency Model} $I_\rho$ is defined in terms of a density operator $\rho_V$ acting on the tensor product of Hilbert spaces labeled by elements of a set $V$.  $I_\rho(U,W|X)$ is true if $U$ and $W$ are quantum conditionally independent given $X$.
\end{Exa}

A \emph{graphoid} is a dependency model that for all disjoint $U,W,X,Y \subseteq V$satisfies the following axioms:
\begin{align}
\text{Symmetry:}\quad 		& I(U,W|X) \Rightarrow I(W,U|X) \\
\text{Decomposition:}\quad 	& I(U,W\cup Y|X) \Rightarrow I(U,W|X) \\
\text{Weak Union:}\quad 		& I(U,W \cup Y|X) \Rightarrow I(U,W|X \cup Y) \\
\text{Contraction:}\quad 		& I(U,W|X) \,\, \text{and} \,\, I(U,Y|X \cup W) \Rightarrow I(U,W \cup Y|X).
\end{align}
A \emph{positive graphoid} is a graphoid that also satisfies the additional axiom
\begin{align}
\text{Intersection:}\quad		& I(U,W |X \cup Y) \,\, \text{and} \,\, I(U,Y|W \cup X) \Rightarrow I(U,W \cup Y|X).
\end{align}

\begin{The}
The quantum dependency model is a graphoid.
\end{The}
\begin{proof}
Symmetry is immediate because $S(U:W|X)$ is invariant under exchange of $U$ and $W$.  Decomposition and Weak Union follow from the strong subadditivity inequality.  Specifically, for $A,B,C \subseteq V$, strong subadditivity asserts that $S(A:B|C) \geq 0$, or in terms of von Neumann entropies
\begin{equation}
\label{Cond:SS}
S(A\cup C) + S(B\cup C) - S(C) - S(A\cup B \cup C) \geq 0.
\end{equation}
Decomposition asserts that if $S(U:W \cup Y|X) = 0$ then $S(U:W|X) = 0$.  This is true if $S(U:W \cup Y|X) - S(U:W|X) \geq 0$, since $S(U:W|X)$ is guaranteed to be positive by strong subadditivity.  Expanding  $S(U:W \cup Y|X) - S(U:W|X)$ and canceling terms gives
\begin{equation}
\begin{split}
S(U:W \cup Y|X) - S(U:W|X) = & S(U \cup W \cup X) + S(W \cup X \cup Y) \\
& - S(W \cup X) - S(U \cup W \cup X \cup Y),
\end{split}
\end{equation}
but the right hand side is positive by eq.~\eqref{Cond:SS} with $A= U, B = Y, C = W\cup X$.

Weak Union is proved via a similar argument applied to $S(U:W \cup Y|X) - S(U:W|X \cup Y)$.  It follows from eq.~\eqref{Cond:SS} by taking $A = U, B = Y, C = X$.  Finally, contraction follows from noting that $S(U:W|X) + S(U:Y|X \cup W) = S(U:W \cup Y|X)$, which is straightforward to show by expanding in terms of von Neumann entropies.
\end{proof}

The well-known analogous result for classical probability distributions follows immediately because classical probability distributions can be represented by density matrices that are diagonal in an orthonormal product basis, and for such states the von Neumann entropies of subsystems are equal to the Shannon entropies of the corresponding marginal distributions.  Additionally, if $P(V)$ is positive for all possible valuations of the variables then the associated dependency model is actually a positive graphoid.  The analogous quantum property would be to require that $\rho_V$ is a strictly positive operator, i.e. it is of full rank, but we have not been able to prove that this property implies intersection.

The undirected graph dependency model is also a positive graphoid.  The proof is straightforward, so it is not given here.  The following theorem is important for the theory of Markov networks.
\begin{The}[Lauritzen \cite{Lau06a}]
\label{Cond:UDGraph}
The undirected graph dependency model is equivalent to the dependency model obtained by setting 
$I\left (U,V - (U\cup n(U))|n(U) \right )$ for all $U \subseteq V$, where $n(U)$ is the set of nearest neighbors of $U$, and demanding closure under the positive graphoid axioms.
\end{The}

The condition $I\left (U,V - (U\cup n(U))|n(U) \right )$ defines the \emph{Local Markov Property} on a graph.  Note that although its closure under the positive graphoid axioms is equivalent to the Global Markov Property, this is not the case for a graphoid that doesn't satisfy intersection \cite{Lau06a}.

\subsection{Quantum Markov Networks}

\label{Graph:HC}

Using the terminology of the previous section, the definition of a classical Markov Network can be conveniently reformulated as a pair $(G,P(V))$, where $G=(V,E)$ is an undirected graph and $P(V)$ is a probability distribution over random variables represented by the vertices, such that the graphoid $I_P$ satisfies the local Markov property with respect to the graph $G$.  The definition of a quantum Markov network can now be obtained by replacing the probabilistic dependency model with a quantum dependency model.

Let $G = (V,E)$ be an undirected graph and suppose that each vertex $v \in V$ is associated with a quantum system, also denoted $v$, with Hilbert space $\mathcal{H}_{v}$.  Let $\rho_V$ be a state on $\mathcal{H}_V = \bigotimes_{v \in V} \mathcal{H}_v$.  $(G,\rho_V)$ is a \emph{Quantum Markov Network} if the graphoid $I_\rho$ satisfies the local Markov property with respect to the graph $G$.  Further, if $\rho_V$ is of full rank, then $(G,\rho_V)$ is called a \emph{Positive Quantum Markov Network}.  Note that unlike in the classical case, we cannot conclude that the global Markov property holds for positive quantum Markov networks because the intersection axiom has not been proved.

The remainder of this section provides some partial characterization results for quantum Markov networks, along the lines of the Hammersley-Clifford theorem.  The most generally applicable of these results makes use of the $\pity$ product.
\begin{The}
\label{Graph:QHC}
Let $G = (V,E)$ be an undirected graph and let $\mathfrak{C}$ be the set of cliques of $G$.  If $(G,\rho_V)$ is a positive quantum Markov network then there exist positive operators $\sigma_C$ acting on the cliques of $G$, i.e. $C \in \mathfrak{C}$, such that
\begin{equation}
\label{Graph:QHCDecomp}
\rho_V = \bigodot_{C \in \mathfrak{C}} \sigma_C .
\end{equation}
\end{The}
This theorem is analogous to one direction of the Hammersley-Clifford theorem and the proof is very similar to a standard proof for the classical case \cite{Pol04a}, but is somewhat involved so it is given in appendix \ref{Proof}.  However, unlike the classical case, the converse does not hold, i.e. there are states of the form eq. \eqref{Graph:QHCDecomp} that do not satisfy the local Markov property as illustrated by the following example.  

\begin{Exa}
\label{ex:heisenberg}
Consider a chain of 3 qubits $A$, $B$, and $C$ coupled through an anti-ferromagnetic Heisenberg interaction $H = \sigma^x_A\sigma^x_BI_C + \sigma^y_A\sigma^y_BI_C + \sigma^z_A\sigma^z_BI_C + I_A\sigma^x_B\sigma^x_C + I_A\sigma^y_B\sigma^y_C + I_A\sigma^z_B\sigma^z_C$ where $\sigma^x$, $\sigma^y$, and $\sigma^z$ denote the Pauli operators
\begin{equation}
\sigma^x= 
\left( \begin{array}{cc}
0 & 1 \\
1 & 0
\end{array} \right),\ \ 
\sigma^z= 
\left( \begin{array}{cc}
1 & 0 \\
0 & -1
\end{array} \right), \ \ 
\mathrm{and}\ \ 
\sigma^y= \sigma^z \sigma^x.
\end{equation}
The Gibbs state $\rho_{A\cup B \cup C}(\beta) = \frac{1}{Z(\beta)} \exp(-\beta H)$ has the form eq. \eqref{Graph:QHCDecomp}, but for any finite $\beta$ it has a non-zero mutual information between $A$ and $C$ conditioned on $B$ as shown on Fig.~\ref{Heisenberg}.
\begin{figure}[h!]
\center \includegraphics[height=2.5in]{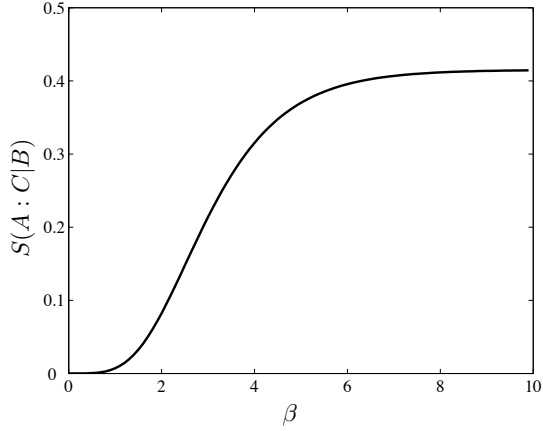}
\caption{Conditional mutual information for a 3-vertex anti-ferromagnetic Heisenberg spin-$\frac 12$ chain as a function of inverse temperature $\beta$.}
\label{Heisenberg}
\end{figure}
\end{Exa}

For trees, a decomposition into reduced and mutual density operators analogous to eq. \eqref{Graph:HCMut} is possible.  For this, we need the following lemma.

\begin{Lem}
\label{Graph:Remove}
Let $G= (V,E)$ be a graph, let $(G,\rho_V)$ be a quantum Markov network and let $u \in V$.  Let $G' = (V',E')$ be the graph obtained by removing $u$ from $V$ and removing all edges that connect $u$ to any other vertex from the graph.  Let $G'' = (V',E'')$ be the graph obtained by adding to $G'$ an edge between every pair of distinct neighbors of $u$ in the original graph $G$.  Let $\rho_{V'} = \PTr{u}{\rho_{V'}}$.  Then $(G'',\rho_{V'})$ is a quantum Markov network.
\end{Lem}
\begin{proof}
For $U \subset V$, let $U_u = U-u$ if $u \in U$ and $U_u = U$ otherwise, and denote $n_G(U_u)$ and $n_{G''}(U_u)$ the neighbors of $U_u$ in the graphs $G$ and $G''$ respectively.  It must be shown that $I_{\rho_V}(U, V-(U \cup n_G(U) | n_G(U))$ for all $U \subset V$ implies $I_{\rho_{V'}}(U_u, V'-(U_u \cup n_{G''}(U_u) | n_{G''}(U_u))$ for every $U_u \subset V'$. By symmetry, we can assume without loss of generality that $u \in U$. There are two different cases to consider:

\noindent{\bf Case I:} $n_G(u) \cap U \neq \emptyset$.\\
This implies that $n_{G''}(U_u) = n_G(U)$ and so $V' - (U_u \cup n_{G''}(U_u)) = V - (U \cup n_G(U))$. We conclude that $I_{\rho_{V'}}(U_u, V'-(U_u \cup n_{G''}(U_u) | n_{G''}(U_u))$ is equivalent to $I_{\rho_V}(U - u, V-(U \cup n_G(U) | n_G(U))$, and the result follows from decomposition.
 
\noindent{\bf Case II:} $n_G(u) \cap U = \emptyset$.\\
This implies that $n_{G''}(U_u) = n_G(U_u)$. Consider the local Markov property on the original graph $G$ applied to $U_u$: $I_{\rho_{V}}(U_u, V-(U_u \cup n_{G}(U_u) | n_{G}(U_u))$ which is equivalent to  $I_{\rho_{V}}(U_u, u \cup V'-(U_u \cup n_{G''}(U_u) | n_{G''}(U_u))$, and the result follows from decomposition.  
\end{proof}

\begin{The}
Let $G= (V,E)$ be a tree.  If $(G,\rho_V)$ is a positive quantum Markov network then it can be written as
\begin{equation}
\rho_V = \left (\bigotimes_{v \in V} \rho_v \right ) \pns{n} \left ( \prod_{(v,u) \in E} \rho\ns{n}_{v:u}   \right ).
\label{eq:standard_graph_state}
\end{equation}
\label{The:mutual}
\end{The}

\begin{proof}
The proof is by induction on the number of vertices in the tree.  It is clearly true for a single vertex, so consider a tree $G=(V,E)$ with $N$ vertices and choose a leaf vertex $u \in V$.  Construct the quantum Markov network $(G'',\rho_{V'})$ as in lemma \ref{Graph:Remove}.  Since $u$ is a leaf it only has one neighbor in $G$, denoted $w$, so the only difference between $G$ and $G''$ is that $u$ and the single edge connecting $u$ to the rest of the graph have been removed.  By the inductive assumption, $\rho_{V'}$ has a  decomposition of the form
\begin{equation}
\label{Graph:InductAssump}
\rho_{V'} = \left (\bigotimes_{v \in V'} \rho_v \right ) \pns{n} \left ( \prod_{(v,x) \in E''} \rho\ns{n}_{v:u}   \right ).
\end{equation} 
Generally, $\rho_{V} = \rho_{V' \cup \{u\}} = \rho_{V'} \pns{n} \rho\ns{n}_{u|V'}$.  The local Markov property implies that $I_\rho (u,V' - w|w)$, so that $\rho_{u|V'} = \rho_{u|w}$, which in turn can be written as $\rho_{u|w} = \rho_u \pns{n} \rho_{u:w}$, so 
\begin{equation}
\rho_V = \rho_{V'} \pns{n} \left ( \rho_u \pns{n} \rho_{u:w}\right  ).
\end{equation}  
Every term in eq.~\eqref{Graph:InductAssump} commutes with $\rho_u$, because they are defined on different tensor product factors.  Also, $\rho_{u:w}$ commutes with all the other mutual density operators either because they act on different tensor product factors or because the fact that $w$ is the only neighbor of $u$ implies that $u$ is quantum conditionally independent of any other subsystem given $w$.
\end{proof}

In the classical case, the Hammersley-Clifford decomposition is not necessarily unique, and when the graph is a tree the decomposition into marginal and mutual distributions is only one possibility. Similarly, a state $\rho_V$ might have a decomposition of the form of eq.~\eqref{eq:standard_graph_state} but with more general operators in place of the mutual and marginal states. This provides another motivation for the definition of an $n$-bifactor state that was given in eq.~\eqref{Graph:QGSE}.   As mentioned in \S\ref{Graph:QGS}, not all $n$-bifactor states are quantum Markov networks, but a subset of them are, as shown by the following theorem.

\begin{The}
\label{Graph:THCC}
Let $G = (V,E)$ be a tree with each vertex $v \in V$ associated to a quantum system with Hilbert space $\mathcal{H}_v$.  Let $\mathcal{H}_V = \bigotimes_{v \in V} \mathcal{H}_v$ and let $\rho_V$ be an $n$-bifactor state on $\mathcal{H}_V$.  If $\mu_v$ is decomposable with respect to all pairs $\nu_{u:v}$ and $\nu_{w:v}$, then $(G, \rho_V)$ is a quantum Markov network.
\end{The}

The notion of decomposability used in the statement of this theorem is defined at eq. \eqref{def:decomposable}. The proof is straightforward and we leave it as an exercise. 

\subsection{Other Graphical Models}

\label{Graph:OM}

In this section quantum generalizations of two other Graphical Models are described: Factor Graphs and Bayesian Networks.  Generally, the choice of which model to use depends on the application and Belief Propagation algorithms have been developed for all of them in the classical case.  For example, Factor Graphs arise naturally in the theory of error correcting codes, Bayesian Networks are commonly used to model causal reasoning in artificial intelligence, and Markov Networks are useful in statistical physics. However, it is now understood that the classical versions of these three models are interconvertable, and that upon such conversion the different Belief Propagation algorithms are all equivalent in complexity \cite{AM00a, YFW02a, KFL01a}.  Some similar results also hold for the quantum case, as we illustrate by showing how a quantum factor graph can be converted into a $1$-Bifactor Network. This construction is used in the application to quantum error correction described in \S\ref{App:QEC}.  

\subsubsection{Quantum Factor Graphs}

\label{Graph:FG}

\begin{figure}
\center\includegraphics{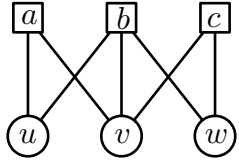}
\caption{Factor graph representation of the state $(\Ket{000} + \Ket{111})_{uvw}$, with $\mu_u = \mu_v = \mu_w = I$ and $X_a = (I+\sigma^z_u\otimes \sigma^z_v)$, $X_b = (I+\sigma^x_u\otimes \sigma^x_v\otimes\sigma^x_w)$, and $X_c = (I+\sigma^z_v\otimes\sigma^z_w)$.}
\label{fig:FG}
\end{figure}

A \emph{quantum factor graph} consists of a pair $(G, \rho_V)$, where $G = (U,E)$ is a bipartite graph and $\rho_V$ is a quantum state.  A bipartite graph is an undirected graph for which the set of vertices can be partitioned into two disjoint sets, $V$ and $F$, such that $(v,f) \in E$ only if $v \in V$ and $f \in F$. The vertices in $V$ are referred to as ``variable nodes" and those in $F$ as ``function nodes".  Each variable node $v$ is associated with a quantum system, also labeled $v$, with a Hilbert space $\mathcal{H}_v$, and $\rho_V$ is a state on $\bigotimes_{v \in V} \mathcal{H}_v$. The Hilbert space associated to a function node $f$ is the tensor product of the Hilbert spaces of the adjacent variable nodes\footnote{The following equality is not just meant in the sense of an isomorphism, they are the same Hilbert spaces.}: $\mathcal{H}_f = \bigotimes_{v \in n(f)} \mathcal{H}_v$. The state associated with a factor graph is of the form
\begin{equation}
\label{Graph:FGE}
\rho_V = \frac 1Z \prod_{f \in F}  X_f \po \bigotimes_{v \in V} \mu_v 
\end{equation}
where $\mu_v$ is an operator on $\mathcal{H}_v$, $X_f$ is an operator on $\mathcal{H}_f$ and $[X_f,X_g] = 0$. 

For example, such a state would be obtained after performing a sequence of projective von Neumann measurements on a product state of the variable nodes (see Fig.~\ref{fig:FG}). More precisely, for each $f \in F$, let $\{P_f^j\}$ be a complete set of orthogonal projectors, and let $\bigotimes _{v \in V} \mu_v$ be the initial state of $V$. When the projective measurements $\{P_f^j\}$ are performed at each function node and commuting outcomes $P_f^j = X_f$ are obtained, the post-measurement state is of the form of eq. \eqref{Graph:FGE}. Similarly, factor graph states could be obtained from more general POVM measurements $\{E_f^j\}$, provided the state update rule $\rho_V \rightarrow \frac{(E_f^j)^{\frac{1}{2}} \rho_V (E_f^j)^{\frac{1}{2}}}{\Tr{E_f^j \rho_V}}$ is used. In that case, the $X_f$ could be any positive operator rather than being restricted to projectors as in the case of a von Neumann measurement. 

To convert a factor graph into a $1$-Bifactor Network, we need to treat the function nodes as distinct quantum systems, and so endow them with their own Hilbert spaces $\mathcal{H}_f = \bigotimes_{v \in n(f)} \mathcal{H}_{R_v^f}$ where $ \mathcal{H}_{R_v^f}$ is isomorphic to $ \mathcal{H}_{v}$. The system $R_v^f$ is called a reference system for $v$ in $f$. Then, the state of the function nodes can be written on the graph $G = (U,E)$, where $U = V\cup F$, $\rho_V = \PTr{F}{\rho_{U}}$ and
\begin{equation}
\rho_U = \frac 1Z  \bigotimes_{u \in U} \mu_u \po \prod_{(v,f) \in E} \nu_{v:f},
\end{equation}
where for $u \in F$, $\mu_u = X_u^T$, $\nu_{v:f} = d_v\kb{\Phi}{\Phi}_{v\cup R_v^f} \otimes I_{f-R^f_v}$ and $\Ket\Phi_{v\cup R_v^f} = \frac{1}{\sqrt d_v} \sum_{j=1}^{d_v} \Ket{j}_v\Ket{j}_{R_v^f}$ denotes the maximally entangled state between $v$ and its reference $R_v^f$.  

\subsubsection{Quantum Bayesian Networks}

\label{Graph:BN}

\begin{figure}
\center\includegraphics{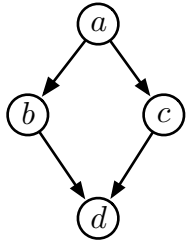}
\caption{This directed acyclic graph has two distinct ancestral orderings: $(a,b,c,d)$ and $(a,c,b,d)$. The equalities $S(d:a|b\cup c) = 0$ and $S(b:d\cup d | a) = 0$ are examples of constraints that are satisfied when $(G,\rho_V)$ is a Quantum Bayesian Network.}
\label{fig:BN}
\end{figure}

Apart from Markov Networks, there are other Graphical Models that make use of the theory of dependency models and graphoids.  Bayesian Networks provide an example, and they are commonly applied in expert systems to model causal reasoning \cite{Nea90a,Nea04a}.  The basic idea is to replace the undirected graph of a Markov network with a Directed Acyclic Graph (DAG), wherein the directed edges represent direct cause-effect relationships.  The quantum graphoid can be used to give a straightforward generalization of the classical networks, which we only treat briefly here.  To describe the generalization, a few definitions and facts about DAGs are required.

For a vertex $v$ in a DAG $G = (V,E)$, let $m(v)$ denote the parents of $v$, i.e. $m(v) = \{u \in V|(u,v) \in E\}$.  The set of ancestors of $v$ is denoted $a(v)$ and consists of those vertices $u$ for which there exists a path in the graph starting at $u$ and ending at $v$.  Conversely, the set of descendants of $v$ is denoted $d(v)$ and consists of those vertices $u$ for which there exists a path in the graph starting at $v$ and ending at $u$.  The set of parents of a subset $U \subseteq V$ of vertices is defined as $m(U) = \cup_{u \in U} m(u) - U$ and similarly $a(U) = \cup_{u \in U} a(u) - U$ and $d(U) = \cup_{u \in U} d(u) - U$.  The set of nondescendants of a subset $U \subseteq V$ of vertices is defined to be $nd(U) = V - (d(U)\cup U)$.  Note that the vertices in $U$ are not considered to be nondescendants of $U$ for technical convenience.  Finally, every DAG has at least one ancestral ordering of its vertices $(v_1,v_2,\ldots,v_n)$, such that if $v_j \in a(v_k)$ then $j < k$ (see Fig. \ref{fig:BN}).

A \emph{Quantum Bayesian Network} is a pair $(G,\rho_V)$, where $G = (V,E)$ is a DAG, each vertex $v \in V$ is associated with a quantum system, also denoted $v$, with Hilbert space $\mathcal{H}_v$, and $\rho_V$ is a quantum state on $\mathcal{H}_V = \bigotimes_{v \in V} \mathcal{H}_v$.  The state $\rho_V$ satisfies the conditional independence constraints $I_\rho (U, nd(U)-m(U)|m(U) )$ for all subsets $U \subseteq V$.

The definition of a classical Bayesian Network is obtained by replacing the quantum systems with classical random variables.  It can be shown that $(G,P(V))$ is a classical Bayesian Network iff $P(V) = \prod_{v \in V} P(v|m(v))$, and a partial quantum generalization of this can be obtained using the conditional density operator.

Due to the nonassociativity of the $\pns{n}$ products, expressions like $A \pns{n} B \pns{n} C$ are ambiguous.  It is convenient to adopt the convention that they are evaluated left-to-right, so that $A \pns{n} B \pns{n} C = \left ( A \pns{n} B \right ) \pns{n} C$.  Similarly, we adopt the convention that
\begin{equation}
\left ( \pns{n} \right )_{j=1}^N A_j = \left ( \left ( \left ( A_1 \pns{n} A_2 \right ) \pns{n} A_3 \right ) \ldots \right ) \pns{n} A_N.
\end{equation}

\begin{The}
If $(G,\rho_V)$ is a Quantum Bayesian Network and $(v_1,v_2,\ldots,v_N)$ is an ancestral ordering of $V$ then
\begin{equation}
\rho_V = \left ( \pns{n} \right )_{j=1}^N \rho\ns{n}_{v_j|m(v_j)}.
\end{equation}
\end{The}
\begin{proof}
For any ordering $(v_1,v_2,\ldots,v_N)$ of the vertices, an arbitrary state can always be written as
\begin{equation}
\rho_V = \left ( \pns{n} \right )_{j=1}^{N} \rho\ns{n}_{v_{j}|v_{j - 1} v_{j - 2} \ldots v_1}.
\end{equation}
This is a quantum generalization of the chain rule for conditional probabilities, which follows straightforwardly from the definition of conditional density operators.  If $(v_1,v_2,\ldots,v_N)$ is in fact an ancestral ordering, then $\{v_{j-1}, v_{j - 2}, \ldots, v_1\} \subseteq nd(v_{j})$, so $I_\rho (v_{j}, nd(v_{j})|m(v_{j}))$ implies that $\rho\ns{n}_{v_j|v_{j-1} v_{j-2} \ldots v_1} = \rho\ns{n}_{v_j|m(v_j)}$.
\end{proof}

% ***** 4) QUANTUM BELIEF PROPAGATION *****

\section{Quantum Belief Propagation}

\label{QBP}

In this section, we discuss algorithms for solving the inference problem that we started with in \S\ref{Problem} for the case of $n$-Bifactor Networks.  In fact, we start with the seemingly simpler problem of computing the reduced density operators of the state on the vertices and on pairs of vertices connected by an edge, and then present a simple modification of the algorithm to solve the inference problem for local measurements.  

Recall that $n$-bifactor states are of the form
\begin{equation}
\label{Graph:QGSERem}
\rho_V = \frac 1 Z \left(\bigotimes_{u\in V} \mu_u\right) \pns{n} \left(\prod_{(v,w) \in E} \nu_{v:w}\right),
\end{equation}
and that the operators associated with vertices and edges do not have to be straightforwardly related to the reduced and mutual density operators.  Therefore, it is not clear a priori that even the simpler task can be done efficiently.  {\em Quantum Belief Propagation} (QBP) algorithms are designed to solve this problem by exploiting the special structure of $n$-bifactor states.  Since the class of states under consideration is different for each value of $n$, there is not one but a family of algorithms. The algorithm that is designed to solve inference problems on $n$-Bifactor Networks is denoted QBP$^{(n)}$.  

To avoid cumbersome notation, focus will be given to $n$-bifactor states with $n < \infty$. Recall that the operators $\nu_{u:v}$ defining these states mutually commute. This is not true of $\infty$-bifactor states. Nevertheless, a Belief Propagation algorithm for $\infty$-bifactor states can be readily defined from the finite $n$ one, by replacing {\em all} products appearing in eqs.~(\ref{message}-\ref{belief2}) by the $\odot$ product. Under this modification, the convergence Theorem \ref{thm:QBP} applies to $\infty$-Bifactor Networks, and its proof only requires straightforward modifications.

The remainder of this section is structured as follows.  \S\ref{QBP:Desc} gives a description of the QBP algorithms and \S\ref{QBP:Conv} shows that QBP$\ns{n}$ converges on trees if the $n$-Bifactor Network is also a quantum Markov Network and that QBP$\ns{1}$ converges on trees in general.  In both cases, the algorithm converges in a time that scales linearly with the diameter of the tree.  Finally, \S\ref{QBP:Inf} explains how to modify the algorithm to solve inference problems for local measurements.

\subsection{Description of the Algorithm}

\label{QBP:Desc}

To describe the operation of the QBP algorithms, it is helpful to imagine that the graph $G$ represents a network of computers with a processor situated at each vertex.  The algorithm could equally well be implemented on a single processor, in which case the network is just a convenient fiction.  Pairs of processors are connected by a communication channel if there is an edge between the corresponding vertices.  The processor at vertex $u$ has a memory that stores the value of $\mu_u$ as well as the value of $\nu_{u:v}$ for each vertex $v$ that is adjacent to $u$ in the graph.  The task assigned to each processor is to compute the local reduced state $\rho_u$ and the joint states $\rho_{u\cup v}$\footnote{Of course, it would be sufficient to only have one processor compute $\rho_{u\cup v}$ for each edge.}.  At each time step $t$, the processor at $u$ updates its ``beliefs'' about $\rho_u$ and $\rho_{u\cup v}$ via an iterative formula.  These beliefs are denoted $b_u\ns{n}(t)$ and $b_{uv}\ns{n}(t)$, and are supposed to be approximations to the true reduced states $\rho_v$ and $\rho_{u\cup v}$ based on the information available to the processor at time step $t$.  Since the reduced states may depend on information stored at other vertices, the processors pass operator valued messages $m\ns{n}_{u \rightarrow v}(t)$ along the edges at each time step in order to help their neighbors.  The message  $m\ns{n}_{u \rightarrow v}(t)$ is an operator on $\mathcal{H}_v$ and is initialized to the identity operator  $m\ns{n}_{u\rightarrow v}(0) = I_v$ at $t = 0$.  For $t > 0$ it is computed via the iterative formula
\begin{equation}
m^{(n)}_{u \rightarrow v}(t) = \frac 1Y \PTr{u}{\mu_u \pns{n} \Bigg[ \Big\{\prod_{v' \in n(u)-v} m^{(n)}_{v' \rightarrow u}(t-1)\Big\} \pns{n} \nu_{u:v} \Bigg]}.
\label{message}
\end{equation}
Here, $Y$ is an arbitrary normalization factor that should be chosen to prevent the the matrix elements of $m\ns{n}_{u \rightarrow v}(t)$ becoming increasingly small as the algorithm proceeds.  It is convenient to choose $Y$ such that $\PTr{v}{m\ns{n}_{u \rightarrow v}(t)} = 1$.

The beliefs about the local density operator $\rho_u$ at time $t$ are given by the simple formula
\begin{equation}
b^{(n)}_u(t)  = \frac 1{Y'} \mu_u \pns{n} \prod_{v' \in n(u)} m^{(n)}_{v' \rightarrow u}(t), \label{belief1}
\end{equation}
where $Y'$ is again a normalization factor that should be chosen to make $\PTr{u}{b^{(n)}_u(t)} = 1$.
On the other hand, the beliefs about $\rho_{u\cup v}$ also depend on the messages received by the processor at $v$, so we have to imagine that each vertex shares its messages with its neighbors.  Having done so, the beliefs about $\rho_{u\cup v}$ are computed via
\begin{equation}
b^{(n)}_{uv}(t) = \frac 1{Y''} (\mu_u \mu_v) \pns{n} \Bigg[\Big\{\prod_{w \in n(u)-v} m^{(n)}_{w \rightarrow u}(t) \prod_{w' \in n(v)-u} m^{(n)}_{w' \rightarrow v}(t)\Big\} \pns{n} \nu_{u:v} \Bigg] \label{belief2},
\end{equation}
where $Y''$ is again a normalization factor.

The beliefs obtained from the QBP$^{(n)}$ algorithm on input $\{\mu_u\}_{u\in V}$ and $\{\nu_{u:v}\}_{(u,v) \in E}$ after $t$ time steps are denoted $[b^{(n)}_{u}(t), b^{(n)}_{uv}(t) ] = \mathrm{QBP}_t^{(n)}(\mu_u,\nu_{u:v})$.  The goal of the next section is to provide conditions under which the beliefs represent the exact solution to the inference problem, i.e. to find states and values of $t$ such that $\mathrm{QBP}_t\ns{n}(\mu_u,\nu_{u:v}) = [\rho_u,\rho_{u\cup v}]$. 
\subsection{Convergence on Trees}

\label{QBP:Conv}

At time $t$, the beliefs $b^{(n)}_{u}(t)$ and $b^{(n)}_{uv}(t)$ represent estimates of the reduced states $\rho_u$ and $\rho_{u\cup v}$ of the input $n$-bifactor state $\rho_V$. 
Note that when the $\mu_u$ and the $\nu_{u:v}$ all commute with one another and are diagonal in local basis, the QBP$^{(n)}$ algorithms all coincide for different $n$ (including $n=\infty$) and correspond to the well known classical Belief Propagation algorithm. This algorithm always converges on trees in a time that scales like the diameter of the tree. Its convergence on general graphs is not fully understood and constitutes an active area of research \cite{Yed01a, YFW02a}. In the quantum setting, the $\mu_u$ and the $\nu_{u:v}$ do not commute in general, but for finite $n$, the $\nu_{u:v}$ commute with each other by assumption. This has straightforward consequence that will be of use later.

\begin{Prop}
\label{prop:commute}
For all $u,v \in V$, $x \in n(u)$, and $w \in n(v)$, the following commutation relations hold $[\nu_{u:v},m^{(n)}_{x\rightarrow u}(t)] = 0$ and $[m^{(n)}_{w\rightarrow v}(t),m^{(n)}_{x\rightarrow u}(t)] = 0$ .
\end{Prop}

Before proving the convergence of Quantum Belief Propagation, the following classical example can help build intuition of its workings, and also serves to outline the crucial steps in proving convergence.

\begin{figure}
\center\includegraphics{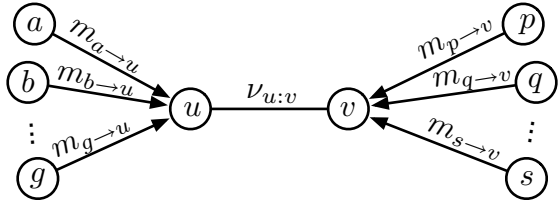}
\caption{Belief $b_{uv}$ is a function of $\mu_u$, $\mu_v$, $\nu_{u:v}$, and the incoming messages at vertices $u$ and $v$, except $m_{u\rightarrow v}$ and $m_{v\rightarrow u}$.}
\label{fig:belief}
\end{figure}

\begin{Exa}
\label{ex:classical}
Consider the function $P$ of $N$ discrete variables $x_j \in \{1,2,\ldots,d\}$ 
\begin{equation}
P(x_1,x_2,\ldots,x_N) = \psi(x_1,x_2)\psi(x_2,x_3)\ldots\psi(x_{N-1},x_N)
\label{eq:classic_ex}
\end{equation}
which could be for instance a classical bifactor distribution on a chain with $N$ sites. To evaluate the marginal function $P(x_N) = \sum_{x_1,x_2,\ldots,x_{N-1}} P(x_1,x_2,\ldots,x_N)$, one can proceed directly and carry the sum over $d^N$ terms. A more efficient solution is obtained by invoking the distributive law to reorder the various sums and products into
\begin{equation*}
P(x_N) = \sum_{x_{N-1}}\Big(\psi(x_{N-1},x_N)\Big( \ldots \Big(\sum_{x_2} \psi(x_2,x_3)\Big(\sum_{x_1} \psi(x_1,x_2) \Big)\Big) \ldots\Big) \Big),
\end{equation*}
and performing the sums sequentially, starting with $\sum_{x_1}$, then $\sum_{x_2}$, and so on
\begin{eqnarray*}
P(x_N) &=& \sum_{x_{N-1}}\Big(\psi(x_{N-1},x_N)\Big( \ldots \Big(\sum_{x_2} \psi(x_2,x_3)M_{1\rightarrow 2}(x_2)\Big) \ldots\Big) \Big) \\
&=& \sum_{x_{N-1}}\Big(\psi(x_{N-1},x_N)\Big( \ldots M_{2\rightarrow 3}(x_3) \ldots\Big) \Big) \\
&&\vdots \\
&=& \sum_{x_{N-1}} \psi(x_{N-1}:x_N) M_{N-2 \rightarrow N-1}(x_{N-1}) 
\end{eqnarray*}
where the ``messages" are defined recursively $M_{j \rightarrow j+1}(x_{j+1}) = \sum_{x_{j}} \psi(x_j:x_{j+1}) M_{j-1 \rightarrow j}(x_j)$, with $M_{1 \rightarrow 2} = \sum_{x_1} \psi(x_1:x_2)$. Each of these steps involves the sum of $d^2$ terms, so $P(x_N)$ can be computed with order $Nd^2$ operations. 
\end{Exa}

This example differs from the Belief Propagation algorithm described in the previous section in three important aspects. Firstly, it relied on the distributive law, which does not hold in general for the $\pns{n}$ product, i.e. $\PTr{u}{X_{uv} \pns{n} Y_{vw}} \neq \PTr{u}{X_{uv}} \pns{n} Y_{vw}$ in general. This will motivate Theorems \ref{thm:commute} and \ref{thm:commute1}, that establish necessary conditions for the validity of the distributive law. Secondly, the graph in that example is a chain, whereas Belief Propagation operates on any graph. However, Belief Propagation is only guaranteed to converge on trees, and the above example generalizes straightforwardly to such graphs. Thirdly, the messages in the example must be computed in a prescribed order: $M_{i-1\rightarrow i}$ is required to compute $M_{i \rightarrow i+1}$. This last point is important and deserves an extensive explanation.  

Suppose that instead of computing the messages $M_{i\rightarrow i+1}$ sequentially, messages at each vertex were computed at every time step, following the rule $m_{i \rightarrow i\pm1}(t,x_{i\pm1}) = \sum_{x_i} m_{i\mp1 \rightarrow i}(t-1,x_i)\psi(x_i:x_{i\pm1})$, as in eq. \eqref{message}, with the initialization $m_{i\pm1 \rightarrow i}(0,x_i) = 1$. Then, one can easily verify that for $t\geq i$, $m_{i \rightarrow i+1}(t,x_{i+1}) = M_{i \rightarrow i+1}(x_{i+1})$. In other words, the messages $m_{i\rightarrow i+1}$ become time independent after a time equal to the distance between vertex $i$ the beginning of the chain. This observation can in fact be generalized as follows.

\begin{figure}
\center\includegraphics{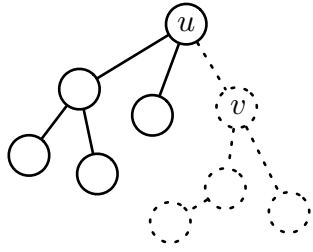}
\caption{For $(u,v) \in E$, the graph $G_v^u$ is obtained from $G$ by considering $u$ as the root and removing the subtree associated to vertex $v$. In this example, $depth(G_v^u) = 2$.}
\label{fig:depth}
\end{figure} 

\begin{Lem}
When $G$ is a tree, the QBP$^{(n)}$ messages $m^{(n)}_{u\rightarrow v}(t)$ are time independent for $t> depth(G_v^u)$, where $G_v^u$ is the tree obtained from $G$ by choosing $u$ as the root, and removing the subtree associated to $v$ (see Fig. \ref{fig:depth}).
\end{Lem}

\begin{proof}
The proof is by induction. If $u$ is a leaf, it has a unique neighbor $n(u)$ and $m_{u \rightarrow n(u)}^{(n)}(t) = \PTr{u}{\mu_u \pns{n} \nu_{u:n(u)}}$ which is time independent. If $u$ is not a leaf, it has two neighbors $L(u)$ and $R(u)$. Clearly, if $m_{L(u)\rightarrow u}^{(n)}(t)$ is time independent for $t \geq t^*$, then $m_{u\rightarrow R(u)}^{(n)}(t) = \PTr{u}{\mu_u \pns{n}\big[m_{L(u)\rightarrow u}^{(n)}(t-1) \pns{n} \nu_{u:R(u)}\big]}$ is time independent for $t \geq t^*+1$. 
\end{proof}

When operated on a tree, all beliefs computed by QBP algorithm converge to a steady state after a time equal to the diameter of the tree. Note that when the graph contains loop, the beliefs do not necessarily reach a steady state. It remains to be shown that on trees, this steady state is the correct solution. For this, we need a technical result that requires some new notation. Let $U$ and $W$ be two non-intersecting subsets of $V$. Define the two subsets of edges $E_U = \{ (u,w) \in E : u,w \in U\}$ and $E_{U:W}= \{ (u,w) \in E : u \in U\ {\mathrm and}\ w \in W\}$.  Let $\Gamma_U = \bigotimes_{u \in U} \mu_u$ and for any $F \subset E$, let $\Lambda_{F} = \prod_{(u,w) \in F} \nu_{v:w}$. 

\begin{The}\label{thm:commute}
Let $(G, \rho_V)$ be an $n$-Bifactor Network with graph $G = (V,E)$. Let $U,W,X$ be non-intersecting subsets of $V$ such that $U\cup W \cup X =V$. When $S(U:X|W) = 0$, the following diagram is commutative.
\begin{equation}\begin{CD}
    \Gamma_{U\cup W} \pns{n} (\Lambda_{E_{U\cup W}} \Lambda_{E_{U\cup W:X}})                 
    @>{\text{Tr}_U}>>           
     \PTr{U}{\Gamma_{U\cup W} \pns{n} (\Lambda_{E_{U\cup W}} \Lambda_{E_{U\cup W:X}})}\\
    @VV{\Gamma_X \pns{n}(\cdot \Lambda_{E_X})}V                      
    @VV{\Gamma_X \pns{n}(\cdot \Lambda_{E_X})}V     \\
    \rho_V = \Gamma_V \pns{n} \Lambda_{E_V} 
    @>{\text{Tr}_U}>> 
    \PTr{U}{\rho_V}
\end{CD}\label{eq:commute_diag}\end{equation}
\end{The}

\begin{proof}
The down-right path is the simplest. The first equality follows from the fact that $\Lambda_{E_X}$ commutes with $\Gamma_{U\cup W}$ and all other $\Lambda_E$'s, and the definition $\rho_V = (\Gamma_{U \cup W} \otimes \Gamma_X )  \pns{n} (\Lambda_{E_{U\cup W}} \Lambda_{E_{U\cup W:X}} \Lambda_{E_X})$. The second equality is just a definition. The right-down path uses the representation of states that saturate strong subadditivity eq.~\eqref{Cond:Hayden}, which implies that $\rho_V$ has a decomposition of the form $\rho_V = \sum^d_{j = 1} p_j \sigma_{U W_j^{(1)}} \otimes \tau_{W_j^{(2)} X}$. First observe that 
\begin{align}
\Gamma_{U \cup W}  \star^{(n)} (\Lambda_{E_{U\cup W}} \Lambda_{E_{U\cup W:X}} ) 
&= (\Gamma_X^{-1}\pns{n} \rho_V) \Lambda_{E_X}^{-1} \\
&= \Big(\Gamma_X^{-1}\pns{n} \sum^d_{j = 1} p_j \sigma_{U W_j^{(1)}} \otimes \tau_{W_j^{(2)} X}\Big)\Lambda_{E_X}^{-1} \\
&=  \sum^d_{j = 1} p_j \sigma_{U W_j^{(1)}} \otimes \Big[\big(\Gamma_X^{-1}\pns{n} \tau_{W_j^{(2)} X}\big)\Lambda_{E_X}^{-1}\Big].
\end{align}
It follows that
\begin{align*}
\Gamma_X\pns{n}\left[\PTr{U}{\Gamma_{U \cup W}  \star^{(n)} (\Lambda_{E_{U\cup W}} \Lambda_{E_{U\cup W:X}} )}\Lambda_{E_X} \right] 
&=  \sum^d_{j = 1} p_j \sigma_{W_j^{(1)}} \otimes \tau_{W_j^{(2)} X}\\
&= \PTr{U}{\rho_V}.
\end{align*}
\end{proof}

Specializing to the case $n=1$ enables a stronger result to be derived that does not require  independence assumptions.

\begin{The}\label{thm:commute1}
Let $(G,\rho_V)$ be a 1-Bifactor Network with graph $G = (V,E)$. Let $U,W,X$ be non-intersecting subsets of $V$ such that $U\cup W \cup X =V$. The following diagram is commutative.
\begin{equation}\begin{CD}
    \Gamma_{U} \po (\Lambda_{E_{U\cup W}} \Lambda_{E_{U\cup W:X}})                 
    @>{\text{Tr}_U}>>           
     \PTr{U}{\Gamma_{U} \po (\Lambda_{E_{U\cup W}} \Lambda_{E_{U\cup W:X}})}\\
    @VV{\Gamma_{W\cup X} \po(\cdot \Lambda_{E_X})}V                      
    @VV{\Gamma_{W\cup X} \po(\cdot \Lambda_{E_X})}V     \\
    \rho_V = \Gamma_V \po \Lambda_{E_V} 
    @>{\text{Tr}_U}>> 
    \PTr{U}{\rho_V}
\end{CD}\end{equation}
\end{The}

\begin{proof}
The theorem follows simply from the cyclic property of the partial trace:
\begin{align}
\PTr{U}{\rho_V}
&= \PTr{U}{[\Gamma_U^\frac 12 \otimes \Gamma_{W\cup X}^\frac 12] \Lambda_{E} [\Gamma_U^\frac 12 \otimes \Gamma_{W\cup X}^\frac 12]} \\
&=  \Gamma_{W\cup X}^\frac 12 \PTr{U}{ \Gamma_U \Lambda_{E}} \otimes \Gamma_{W\cup X}^\frac 12 \\
&=  \Gamma_{W\cup X}^\frac 12 \PTr{U}{ \Gamma_U \Lambda_{E_{U\cup W}} \Lambda_{E_{U\cup W:X}}} \Lambda_{E_X} \otimes \Gamma_{W\cup X}^\frac 12.
\end{align}
\end{proof}

We are now positioned to state and prove the main result of this section. 

\begin{The}
\label{thm:QBP}
Let $(G,\rho_V)$ be an $n$-Bifactor Network with graph $G = (V,E)$, and let $[b^{(n)}_{u}(t), b^{(n)}_{uv}(t) ] = \mathrm{QBP}_t^{(n)}(\mu_u,\nu_{u:v})$. If $(G,\rho_V)$ is a quantum Markov network and $G$ is a tree, then for all $t \geq diameter(G)$, $b^{(n)}_u(t) = \rho_u$ and  $b^{(n)}_{uv}(t) = \rho_{u\cup v}$.
\end{The}

\begin{proof}
First, observe that $b^{(n)}_u(t) = \PTr{v}{b^{(n)}_{uv}(t)}$, so it is sufficient to prove that $b^{(n)}_{uv}(t) = \rho_{u\cup v}$. Consider $u \cup v$ to be the root of the tree. We proceed by induction, repeatedly tracing out leaves from the bifactor state except $u$ and $v$ until we are left with only vertices $u$ and $v$. Set $G(0) = G$ and let $G(t) = (V(t),E(t))$ be the tree left after $t$ such rounds of removing leaves. Denote the leaves of $G(t)$ apart from $u$ and $v$ by $l(t)$, the children of $x$ by $c(x)$, and the unique parent of $x$ by $m(x)$. At $t=0$, consider tracing out a leaf $w$ of $G$
\begin{align}
\PTr{w}{\rho_V}
&= \PTr{u}{(\mu_w\otimes \Gamma_{V-w})\pns{n} (\nu_{w:m(w)} \Lambda_{E_{V-w}})}\\
&= \Gamma_{V-w} \pns{n}\left[\PTr{w}{\mu_w \pns{n} \nu_{w:m(w)}}\Lambda_{E_{V-w}}\right] \\
&= \Gamma_{V-w} \pns{n}\left[m_{w\rightarrow m(w)}^{(n)}(1)\Lambda_{E_{V-w}}\right]
\end{align}
where we have used Theorem~\ref{thm:commute} going from the first to the second line.  Since this holds for all leaves, we conclude that 
\begin{equation}
\PTr{l(0)}{\rho_{V}} = \Gamma_{V(1)} \pns{n} \left(\prod_{x \in l(0)} \prod_{y \in c(x)} m_{y\rightarrow x}^{(n)}(1)\Lambda_{V(1)}\right).
\end{equation}
We thus make the inductive assumption that
\begin{equation}
\label{Graph:IndAss}
\rho_{V(t)} = \Gamma_{V(t)} \pns{n} \left(\prod_{x \in l(t)} \prod_{y \in c(x)} m_{y\rightarrow x}^{(n)}(t)\Lambda_{V(t)}\right).
\end{equation}
It follows that
\begin{align}
\rho_{V(t+1)} 
&= \PTr{l(t)}{\rho_{V(t)}} \\
&= \PTr{l(t)}{\Gamma_{V(t)} \pns{n} \left[\prod_{x \in l(t)} \prod_{y \in c(x)} m_{y\rightarrow x}^{(n)}(t)\Lambda_{V(t)}\right]} \\
&= \PTr{l(t)}{\Gamma_{V(t+1)} \pns{n} \left[\prod_{x \in l(t)}  \mu_x \pns{n}\Big( \prod_{y \in c(x)} m_{y\rightarrow x}^{(n)}(t) \nu_{x:m(x)} \Lambda_{V(t+1)}\Big)\right]} \\
&= \Gamma_{V(t+1)} \pns{n} \left[\prod_{x \in l(t)} \PTr{x}{ \mu_x \pns{n}\Big( \prod_{y \in c(x)} m_{y\rightarrow x}^{(n)}(t) \nu_{x:m(x)}\Big)} \Lambda_{V(t+1)}\right] \\
&= \Gamma_{V(t+1)} \pns{n} \left[\prod_{x \in l(t)} m_{x\rightarrow m(x)}^{(n)}(t+1) \Lambda_{V(t+1)}\right] \\
&= \Gamma_{V(t+1)} \pns{n} \left[\prod_{x \in l(t+1)} \prod_{y \in c(x)} m_{y\rightarrow x}^{(n)}(t+1)\Lambda_{V(t+1)}\right]
\end{align}
also assumes the same form, so eq.~\eqref{Graph:IndAss} follows by induction.  We have again used Theorem~\ref{thm:commute} in going from the third to the fourth line.  When $V(t)$ contains only $u$ and $v$ then this reduces to $\rho_{u\cup v} = b\ns{n}_{uv}(t)$, which is what we set out to prove. 
\end{proof}

Once again, specializing to the case $n=1$ enables a stronger result to be derived that does not rely on independence assumptions.

\begin{Cor}
\label{cor:QBP1}
Let $(G,\rho_V)$ be an $1$-Bifactor Network with graph $G = (V,E)$, and let $[b_{u}(t), b_{uv}(t) ] = \mathrm{QBP}^{(1)}_t(\mu_u,\nu_{u:v})$. If $G$ is a tree, then for all $t \geq diameter(G)$, $b_u(t) = \rho_u$ and  $b_{uv}(t) = \rho_{u\cup v}$.
\end{Cor}

\begin{proof} This Corollary is a consequence of Theorem~\ref{thm:commute1} and the fact that the proof of Theorem~\ref{thm:QBP} only relies on the commutativity of the diagram eq.~\eqref{eq:commute_diag}.
\end{proof}

This last result gives us additional information about the structure of correlations in 1-bifactor states that is captured by the following corollary.

\begin{Cor}
Let $(G,\rho_V)$ be an $1$-Bifactor Network on graph $G = (V,E)$. If $G$ is a tree, then the mutual density operators commute: $[\rho_{u:v},\rho_{w:x}] = 0$ for all $(u,v)$ and $(w,x) \in E$.
\end{Cor}

\begin{proof}
The only non trivial case is $[\rho_{u:v},\rho_{v:w}]$ with $u \neq w$. Let $[b_{u}(t), b_{uv}(t) ] = \mathrm{QBP}^{(1)}_t(\mu_u,\nu_{u:v})$ and denote
\begin{equation}
A_{u-v}(t) = \prod_{w\in n(u)-v} m_{w\rightarrow u}(t).
\end{equation}
Observe that $A_{u-v}(t)$ is an operator on $\mathcal{H}_u$, and by Proposition~\ref{prop:commute}, $[A_{u-v}(t),\nu_{u:w}] = 0$ for all $u$, $v$, and $w\in V$. From Theorem~\ref{thm:QBP}, we have for $t \geq diameter(G)$
\begin{equation}
[\rho_{u:v},\rho_{v:w}] = 
[ A_{u-v}(t) A_{v-u}(t) \nu_{u:v}, A_{v-w}(t) A_{w-v}(t) \nu_{v:w} ] = 0.
\end{equation}
\end{proof}

Corollary~\ref{cor:QBP1} shows that for 1-bifactor states on trees, QBP$^{(1)}$ enables an efficient evaluation of  the one-vertex and two-vertex reduced density operators $\rho_{u}$ for all $u \in V$ and $\rho_{u\cup v}$ for all $(u,v) \in E$. Can this result be generalized to arbitrary bifactor states? This question is of interest since, as we will detail in \S\ref{App:Stat}, the Gibbs states used in statistical physics are $\infty$-bifactor states. However, it is known that approximating the ground state energy of a two-local Hamiltonian on a chain is QMA-complete \cite{AGK07a,Ira07a}\footnote{QMA stands for Quantum Merlin and Arthur and it is the natural quantum generalization of the classical complexity class NP. So to the best of our knowledge, solving a QMA-complete problem would require an exponential amount of time even on a quantum computer.}. Knowledge of $\rho_{u\cup v}$ leads to an efficient evaluation of the energy. Therefore, without any independence assumptions, it is unlikely that an efficient QBP algorithm for $n$-Bifactor Networks will converge to the correct marginals for $n>1$. This contrasts with classical BP that always converges to the exact solution on trees. However, \S\ref{Heur:Rep} gives a QBP algorithm that solves the inference problem for any $n$-bifactor state on a tree in a time that scales exponentially with $n$. 

\subsection{Solving Inference Problems}

\label{QBP:Inf}

We close this section with a discussion of how QBP algorithm can solve inference problems when local measurements are executed on a bifactor state. In other words, for an outcome of a local measurement on a subsystem $U$ described by a POVM element $E^{(j)}_U = \bigotimes_{u \in U} E_u^{(j)}$, we are interested in evaluating the  marginal states $\rho_{u|E^{(j)}_U}$ and $\rho_{u\cup v|E^{(j)}_U}$ conditioned on the outcome, where
\begin{align}
\rho_{u|E^{(j)}_U} & = \frac 1 Y \PTr{V - u}{(E^{(j)}_U)^{\frac{1}{2}} \rho_V (E^{(j)}_U)^{\frac{1}{2}}} \\
\rho_{u\cup v|E^{(j)}_U} & = \frac 1 Y \PTr{V - \{u,v\}}{(E^{(j)}_U)^{\frac{1}{2}} \rho_V (E^{(j)}_U)^{\frac{1}{2}}},
\end{align}
and $Y$ is a normalization factor.
For $u,v \notin U$, this amounts to a local modification of the bifactor state that accounts for the action of the measurement, the QBP algorithm being otherwise unaltered. We focus on 1-Bifactor Networks and return to the general case at the end of this section.

\begin{The}
\label{thm:update}
Let $(G,\rho_V)$ be a 1-Bifactor Network with $G = (V,E)$ a tree. For $U \subset V$, let $\{E^{(j)}_U\} = \Big\{\bigotimes_{u \in U} E_u^{(j)}\Big\}$ be a POVM on the subsystem $U$ and let $W = V-U$. Define $\mu_u^{(j)}  = \mu_u \po E_u^{(j)} $ for $u\in U$ and $\mu_u^{(j)} = \mu_u$ for $u \in W$. Let $[b_{uv}(t), b_{uv}(t) ] = \mathrm{QBP}^{(1)}(\mu_u^{(j)},\nu_{u:v})$. Then for all $t \geq diameter(G)$, $b_u(t) = \rho_{u|E_U^{(j)}}$ for all $u \in W$ and  $b_{u\cup v}(t) = \rho_{uv|E_U^{(j)}}$ for all $(u,v) \in E_W$.
\end{The}  

\begin{proof}
The reduced state on $W$ conditioned on the measurement outcome $E_U^{(j)}$ is given by
\begin{align}
\rho_{W|E_U^{(j)}} & = \frac 1Y \PTr{U}{(E_U^{(j)})^{\frac{1}{2}} \rho_V (E_U^{(j)})^{\frac{1}{2}}} \\
&= \frac 1Y \prod_{\substack{v \in W \\ u\in U}} \prod_{(w,x) \in E} \mu_v^\frac 12  \PTr{U}{(E_u^{(j)})^{\frac{1}{2}} \mu_u^\frac 12 \nu_{w:x} \mu_u^\frac 12 (E_u^{(j)})^{\frac{1}{2}}}\mu_v^\frac 12 \\
&= \frac 1Y \prod_{\substack{v \in W \\ u\in U}} \prod_{(w,x) \in E} \mu_v^\frac 12  \PTr{U}{ \nu_{w:x} \mu_u^\frac 12 E_u^{(j)} \mu_u^\frac 12 }\mu_v^\frac 12 \\
&= \frac 1Y \prod_{\substack{v \in W \\ u\in U}} \prod_{(w,x) \in E} \Big(\mu_v^{(j)}\Big)^\frac 12  \PTr{U}{ \Big(\mu_u^{(j)}\Big)^\frac 12 \nu_{w:x} \Big(\mu_u^{(j)}\Big)^\frac 12 }\Big(\mu_v^{(j)}\Big)^\frac 12 .
\end{align}
The result thus follows from Corollary~\ref{cor:QBP1}.
\end{proof}

The result of Theorem~\ref{thm:update} can easily be extended to compute the conditional marginal state $\rho_{u|E^{(j)}_U}$ and $\rho_{u\cup v|E^{(j)}_U}$ for any $u$ and $v$, not just those in $W = V-U$. This is achieved by altering the beliefs as follows
\begin{equation}
b_u(t)   = \frac 1ZE_u^{(j)} \po \mu_u \po  \prod_{v' \in n(u)} m_{v' \rightarrow u}(t)
\end{equation}
for $u \in U$, 
\begin{equation}
b_{uv}(t) = \frac 1Z E_{uv}^{(j)} \po (\mu_u \mu_v)\po  \Bigg[\prod_{w \in n(u)-v} m_{w \rightarrow u}(t) \prod_{w' \in n(v)-u} m_{w' \rightarrow v}(t) \po \nu_{u:v} \Bigg]
\end{equation}
with $E_{uv}^{(j)} = E_{u}^{(j)} \otimes I_v$ when $u\in U$ and $v \in W$ and $E_{uv}^{(j)} = E_{u}^{(j)} \otimes E_{u}^{(j)} $ when $u,v \in U$. The proof is straightforward and we omit it.

Theorem~\ref{thm:update} shows how QBP leads to an efficient algorithm for solving inference problems on 1-bifactor states on trees with local measurements. This immediately implies an efficient algorithm for general $n$-bifactor states when $(G,\rho_V)$ is a quantum Markov network. Indeed, Theorem~\ref{thm:QBP} demonstrates that in that case the QBP$^{(n)}$ algorithm can be used to efficiently compute the marginal density operators $\rho_{u\cup v}$ for all $(u,v) \in E$. From these, one can straightforwardly obtain the marginal operators $\rho_u$ for all $u \in V$ and mutual operators $\rho_{u:v}$ for all $(u,v) \in E$. Theorem~\ref{The:mutual} states that $\rho_V$ can be represented as a 1-bifactor state in terms of its marginal and mutual operators. The inference problem can then be solved using the QBP$^{(1)}$ algorithm as explained above. 

% ***** 5) HEURISTIC METHODS *****

\section{Heuristic Methods}

\label{Heur}

The previous section provided conditions under which QBP algorithms give exact solutions to inference problems on $n$-Bifactor Networks. Namely, the underlying graph must be a tree, and the state must be either a quantum Markov network or a 1-bifactor state. When these conditions are not met, QBP algorithms may still be used as heuristic methods to obtain approximate solutions to the inference problem, although in general these approximations will be uncontrolled. 

To draw a parallel, classical Belief Propagation algorithms have found applications in numerous distinct scientific fields where they are sometimes known under different name: Gallager decoding, Viterbi's algorithm, sum-product, and iterative turbo decoding in information theory; cavity method and the Bethe-Peierls approximation in statistical physics; junction-tree and Shafer-Shenoy algorithm in machine learning to name a few. In many of these examples, BP algorithms exhibit good performance on graphs with loops, even though the algorithm does not converge to the exact solution on such graphs. In fact,  ``Loopy Belief Propagation" is often the best known heuristic method to find approximate solutions to hard problems. Important examples include the near-Shannon capacity achieving turbo-codes and low density parity check codes. On the other hand, there are known examples for which loopy BP fail to converge and their general realm of applicability is not yet fully understood. 

As in the classical case, one can expect loopy QBP to give reasonable approximations in some circumstances, for instance when the size of typical loops is very large. Intuitively, one expects a local algorithm to be relatively insensitive to the large scale structure of the underlying graph. However, quantum inference problems also pose a new challenge. Quite apart from issues regarding the graph's topology, an $n$-bifactor state with $n>1$ may not obey the independence conditions required to ensure the convergence of QBP. The goal of this section is to suggest three techniques that are expected to improve the performance of QBP in such circumstances. 

\subsection{Coarse-graining}

\label{Heur:CG}

By definition, a quantum Markov network has the property that the correlations from one vertex to the rest of the graph are screened off by its neighbors. When this property fails, QBP will not in general produce the correct solution to an inference problem. Coarse graining is a simple way of modifying a graph in such a way that the state may be a closer to forming a quantum Markov network with respect to the new graph than it was with respect to the original graph. 

\begin{figure}[h!]
\center \includegraphics[height=1.6in]{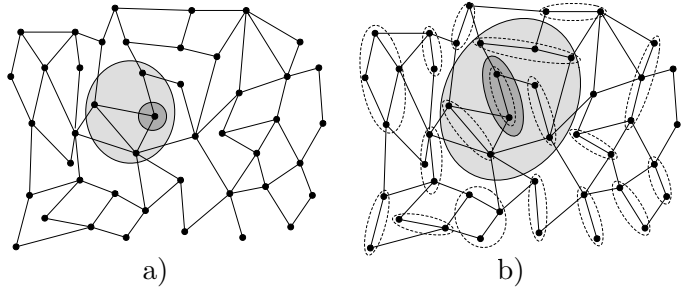}
\caption{Example of a coarse-grained graph. Figure a) shows in light gray the neighborhood of the darkened vertex in the original graph. In b) the dashed ellipses represent coarse-grained vertices. The neighborhood of the darkened coarse-grained vertex is represented by the light gray set.}
\label{CG}
\end{figure}

A coarse graining of a graph $G = (V,E)$ is a graph  $\tilde G = (\tilde V, \tilde E)$, where $\tilde V$ is a partition of $V$ into disjoint subsets of and $(U,W) \in \tilde{E}$ if there is an edge connecting a vertex in $U$ to a vertex in $W$ in $G$.  The coarse-grainings that are of most interest are those that partition $V$ into connected sets of vertices (see Fig.~\ref{CG} for example).  It is an elementary exercise to show that if $(G,\rho_V)$ is an $n$-Bifactor Network, then $(\tilde G, \rho_{\tilde{V}})$ is an $n$-Bifactor Network for any coarse graining $\tilde G$. The intuition for why coarse graining might get us closer to a Markov network is that it effectively ``thickens" the neighborhood of each vertex, which may then be more efficient at screening off correlations. This intuition is illustrated in Fig.~\ref{CG} and is supported by the fact that Markov networks are fixed points of the coarse graining procedure, i.e. if $\tilde G$ is a coarse-graining of $G$, then $(\tilde G, \rho_{\tilde{V}})$ is a quantum Markov network whenever $(G,\rho_V)$ is a Markov network.

Also note that every graph $G$ can be turned into a tree by a suitable coarse graining. When the obtained Bifactor Network is a Markov Network or when $n=1$, QBP is then guaranteed to converge to the exact solution. The Hilbert space dimension at the vertices of the coarse-grained graph is bounded by an exponential in the tree-width of $G$, so this technique is efficient only for graph of $O(\log(N))$ tree-width. 

\subsection{Sliding window QBP}

\label{Heur:SW}

Sliding window QBP is similar in spirit to coarse-graining but is mainly suitable for chains (although the idea is easily generalized to arbitrary trees of low degree). Consider an $n$-bifactor state $\rho_V$ on a one dimensional lattice $G = (V,E)$ with $V = \{v_1,v_2,\ldots,v_N\}$ and $E = \{(v_j,v_{j+1})\}_{j=1,\ldots ,N-1}$. When $(G,\rho_V)$ is not a quantum Markov Network, the diagram of eq.~\eqref{eq:commute_diag} will generally fail to be commutative. The commutativity of this diagram is essential for the success of QBP, as for instance it implies
\begin{align}
\PTr{v_1}{(\mu_{v_1} \otimes \mu_{v_2}) \pns{n} (\nu_{v_1:v_2} \nu_{v_2:v_3})} &= \mu_{v_2} \pns{n}\left[ \PTr{v_1}{\mu_{v_1} \pns{n} \nu_{v_1:v_2}} \pns{n} \nu_{v_2:v_3}\right] \\
&= \mu_{v_2} \pns{n} \big[ m_{v_1\rightarrow v_2} \pns{n} \nu_{v_2:v_3} \big].
\end{align}
Thus, the Hilbert space of vertex $v_1$ is traced out before operators on vertex $v_3$ are brought into the picture. This enables the algorithm to progress along the lattice by evaluating a cumulative operator of constant dimension (i.e. the messages), much in the spirit of the transfer matrix of statistical physics. Without the Markov property, this is generally not possible. 

However, when vertices separated by a distance $\ell$ are conditionally independent given the vertices between them, sliding window QBP can be operated efficiently to produce the exact solution of the inference problem. This works by defining new message operators
\begin{equation}
\tilde m_{v_{j +\ell-1}\rightarrow v_{j+\ell}} = \PTr{\{v_1,v_2, \ldots ,v_j\}}{\left[\bigotimes_{k = 1}^{\ell + j-1} \mu_{v_k}\right] 
\pns{n} \left[\prod_{k = 1}^{\ell+j-1} \nu_{v_k:v_{k+1}}\right]}
\end{equation}
which act on $\mathcal{H}_{v_{j+1}} \otimes \mathcal{H}_{v_{j+2}} \otimes \ldots \mathcal{H}_{v_{j+\ell}}$. When
\begin{equation}
S(v_j:v_{j+\ell}|\{v_{j+1}, v_{j+2}, \ldots , v_{j+\ell-1}\}) = 0
\label{correlation-length}
\end{equation}
for all $v_j \in V$, we have the equality
\begin{equation}
\tilde m_{v_{j+\ell} \rightarrow v_{j+\ell+1}} = \PTr{v_{j+1}}{\mu_{v_{j+\ell}}\pns{n}\big[\tilde m_{v_{j+\ell-1}\rightarrow v_{j+\ell}} \pns{n} \nu_{v_{j+\ell}:v_{j+\ell+1}}\big]},
\end{equation}
so inference problems can be solved exactly with operators whose dimension grow exponentially with the $\ell$ rather than the lattice size $N$.   In particular, this method can be applied to spin-systems that have a finite correlation length because then eq.~\eqref{correlation-length} can be expected to hold approximately for some finite $\ell$.

\subsection{Replicas}

\label{Heur:Rep}

The method of replicas maps $n$-bifactor states to 1-bifactor states on which QBP$^{(1)}$ can be implemented without concerns for independence. This is achieved by replacing the systems $v$ on each vertex of the graph $G$ by $n$ replicas, so that the Hilbert space associated to vertex $v$ becomes $\mathcal{H}_v^{\otimes n}$. As a consequence, the algorithm suffers an overhead exponential in $n$. The name ``replica" is borrowed from the analogous technique used in the study of classical quenched disordered systems. The validity of this technique is based on the following observation. 

\begin{Prop}\label{prop:product}
Let $\{\mathcal{H}_j\}_{j=1,\ldots,n}$ be isomorphic Hilbert spaces. Let $T^{(n)}$ be the operator that cyclicly permutes these $n$ systems. Let $A_1$ be an arbitrary operator on $\mathcal{H}_1$, and define $A_j = (T^{(n)})^{j-1} A_1 (T^{(n)\dagger})^{j-1}$ to be the corresponding operators on $\mathcal{H}_j$. Then for any set of operators $\{A_1^{(k)}\}$ on $\mathcal{H}_1$, the following equality holds
\begin{equation}
A_1^{(1)}A_1^{(2)}\ldots A_1^{(n)} = \PTr{2,3,\ldots, n}{[A_1^{(1)}\otimes A_2^{(2)}\otimes\ldots\otimes A_n^{(n)}]T^{(n)}}.
\end{equation}
\end{Prop}

We are now in a position to formalize the replica method. 

\begin{The} Let $(G,\rho_V)$ be an $n$-Bifactor Network, with operators $\mu_u$ and $\nu_{u:v}$. Then, $\rho_V$ is locally isomorphic to a 1-bifactor state with Hilbert spaces comprising $n$ replicas of the original system $\mathcal{H}_u' = \mathcal{H}_{u_1} \otimes \mathcal{H}_{u_2} \otimes \ldots \otimes \mathcal{H}_{u_n}$ for all $u \in V$. The partial isomorphism at vertex $u$ is given by $\PTr{u_2,u_3,\ldots,u_n}{(T^{(n)\dagger}_u)^\frac 12 \cdot (T^{(n)}_u)^\frac 12}$.  More precisely, we claim that
\begin{equation}
\rho_V = \PTr{\{u_2,u_3,\ldots, u_n\}_{u \in V}}{U^\dagger \left(\bigotimes_{u\in V} \tilde \mu_u\right) \po \left(\prod_{(v,w) \in E} \tilde\nu_{v:w}\right) U}
\end{equation}
where
\begin{align}
\tilde \mu_u &= \left(\mu_u^{\frac{1}{n}} \right)^{\otimes n} \left(T_u^{(n)}\right) \\
\tilde \nu_{u:v} &= \left(\nu_{u:v}^{\frac 1n}\right)^{\otimes n}  \\
U &= \bigotimes_{u \in V} (T^{(n)}_u)^\frac 12
\end{align}
are operators on $\mathcal{H}_u'$
\end{The}

\begin{proof}
First, note that $T_u^{(n)}$ commutes with $\left(\mu_u^{\frac{1}{n}} \right)^{\otimes n}$, so  $\tilde \mu_u^\frac 12 = \left(\mu_u^{\frac{1}{2n}} \right)^{\otimes n} \left(T_u^{(n)}\right)^\frac 12 = \left(T_u^{(n)}\right)^\frac 12 \left(\mu_u^{\frac{1}{2n}} \right)^{\otimes n}$. Thus
\begin{align}
& \PTr{\{u_2,u_3,\ldots, u_n\}_{u \in V}}{U^\dagger \left(\bigotimes_{u\in V} \tilde \mu_u\right) \po \left(\prod_{(v,w) \in E} \tilde\nu_{v:w}\right)U} \\
&= \PTr{\{u_2,u_3,\ldots, u_n\}_{u \in V}}{U^\dagger \left(\bigotimes_{u\in V} T_u^{(n)} \left(\mu_u^{\frac 1n} \right)^{\otimes n}\right) \po \left(\prod_{(v,w) \in E}  \left(\nu_{u:v}^{\frac 1n}\right)^{\otimes n} \right)U} \\
&= \PTr{\{u_2,u_3,\ldots, u_n\}_{u \in V}}{\left(\bigotimes_{u\in V} \left(\mu_u^{\frac 1n} \right)^{\otimes n}\right) \po \left(\prod_{(v,w) \in E}  \left(\nu_{u:v}^{\frac 1n}\right)^{\otimes n} \right)  \bigotimes_{u \in V} T_u^{(n)} } \\
&= \PTr{\{u_2,u_3,\ldots, u_n\}_{u \in V}}{\left[\left(\bigotimes_{u\in V} \mu_u^{\frac 1n} \right) \po \left(\prod_{(v,w) \in E}  \nu_{u:v}^{\frac 1n} \right)\right]^{\otimes n}  \bigotimes_{u \in V} T_u^{(n)} } \\
&= \left[\left(\bigotimes_{u\in V} \mu_u^{\frac 1n} \right) \po \left(\prod_{(v,w) \in E}  \nu_{u:v}^{\frac 1n} \right)\right]^n = \rho_V
\end{align}
where we used Proposition~\ref{prop:product} to obtain the last line. 
\end{proof}

Since the dimension of the Hilbert at each vertex grows exponentially with $n$, the QBP$^{(1)}$ algorithm used to solve the corresponding inference problem suffers an exponential overhead. One can make a replica symmetry ansatz, assuming that the state is symmetric under exchange of replica systems at any given vertex. Since the symmetric subspace of $\mathcal{H}_v^{\otimes n}$ grows polynomially\footnote{More precisely, it grows as $\binom{n+d-1}{n} \approx n^{d-1}$.} with $n$, QBP algorithm can be executed efficiently. The validity of this ansatz cannot be verified in general, but it may serve as a good heuristic method. 

% ***** 5) APPLICATIONS *****

\section{Applications}

\label{App}

This section explains in some detail how QBP can be used as a heuristic algorithm to find approximate solutions to important problems in quantum error correction and the simulation of many-body quantum systems. The focus will be on the reduction of well established problems to inference problems on $n$-Bifactor Networks. One can make use of the techniques discussed in the previous section whenever the resulting Graphical Model does not meet the requirements to ensure convergence of QBP, or when these conditions cannot be verified efficiently.

\subsection{Quantum Error Correction}

\label{App:QEC}

Maximum-likelihood decoding is an important task in quantum error correction (QEC). As in classical error correction, this problem reduces to the evaluation of marginals on a factor graph, also called Tanner graph in this context. More precisely, for independent error models, the quantum channel conditioned on error syndrome is a 1-bifactor state. As a consequence, qubit-wise maximum likelihood decoding of a QEC stabilizer code reduces to an inference problem on a 1-Bifactor Network. Thus, there is no independence condition that needs to be verified, although the graph will generally contain loops. Before demonstrating this reduction, a brief summary of stabilizer QEC is in order, see \cite{Got97a} for more details. For details on the use of Belief Propagation for the decoding of classical error correction codes, the reader is referred to the text of MacKay \cite{Mac03a} and forthcoming book of Richardson and Urbanke \cite{RU05a}.  

Consider a collection of $N$ two-dimensional quantum systems (qubits) $V = \{u\}_{u=1,\ldots,N}$ with $\mathcal{H}_u = \mathbb{C}^2$. A QEC code is a subspace $\mathcal{C} \in \mathcal{H}_V$ that is the $+1$ eigensubspace of a collection of commuting operators $S_j$, $j=1,\ldots N-K$, called stabilizer generators. Each stabilizer generator is a tensor product of Pauli operators on a subset $U_j$ of $V$:
\begin{equation}
S_j = \bigotimes_{u \in U_j} \sigma^{\alpha^u_j}_u
\end{equation}
where $\alpha_j^u \in \{x,y,z\}$. When the stabilizer generators are multiplicatively independent, the code encodes $K$ qubits, i.e. $\mathcal{C}$ has dimension $2^K$. For each $j=1,\ldots N-K$, define the two projectors $P_j^\pm = (I \pm S_j)/2$. The code space is therefore defined as $\mathcal C = (\prod_j P_j^+) \mathcal H_V$. 

Error correction consists of three steps. First, the system $V$ is prepared in a code state $\rho_V$ supported on $\mathcal C$, in such a way that $P_j^+ \rho_V P_j^+ = \rho_V$ for all $j$. The state is then subjected to the channel $\rho_V \rightarrow \mathcal E_{V|V}(\rho_V)$. Second, each stabilizer generator $S_j$ is measured, yielding an outcome $s_j = \pm$ with probability $\Tr{P^\pm_j \mathcal{E}_{V|V}(\rho_V)}$. The collection of all $N-K$ measurement outcomes $s_j$, called the error syndrome, is denoted  ${\bf s} = (s_1,s_2,\ldots s_{N-K})\in \{-,+\}^{N-K}$. Third, the channel $\mathcal E_{V|V}$ is updated conditioned the error syndrome $\bf s$. Based on this updated channel, the optimal recovery is computed and implemented. 

The computationally difficult step in the above protocol consists in conditioning the channel on the error syndrome. To understand this problem, it is useful to express the channel in a Kraus form $\mathcal{E}_{V|V}(\rho_V) = \sum_k M^{(k)}_{V|V} \rho_V M^{(k)\dagger}_{V|V}$ where $\{M^{(k)}\}$ are operators on $\mathcal{H}_V$. When $s_j = +$, we learn that the error that has affected the state commutes with $S_j$, while $s_j = -$ indicates that the error anti-commutes with $S_j$. To update the channel conditioned on the error syndrome $s_j = +$ say, we first decompose each Kraus operator $M^{(k)}_{V|V}$ as the sum of an operator that commutes with $S_j$ and an operator that does not commute with $S_j$: $M^{(k)}_{V|V} = M^{(k)+}_{V|V} + M^{(k)\prime}_{V|V}$ where $ M^{(k)+}_{V|V} = P_j^+  M^{(k)}_{V|V} P_j^+$ and $ M^{(k)\prime}_{V|V} =  M^{(k)}_{V|V} -  M^{(k)+}_{V|V}$. The updated channel is obtained by throwing away the primed component  $ M^{(k)\prime}_{V|V}$ of each Kraus operator, and renormalizing.  

In what follows, we demonstrate how the conditional channel can be expressed as a factor graph. This is most easily done using the Jamio\l kowski representation of quantum channels. For each quantum system $v$, let $R_v$ denote a reference for $v$, with Hilbert space $\mathcal{H}_{R_v} \simeq \mathcal H_v$. Define the maximally entangled state between system $v$ and its reference by $\Ket\Phi_{vR_v} =  \frac{1}{\sqrt d} \sum_j \Ket{j}_v\Ket{j}_{R_v}$. Then, the Jamio\l kowski representation of a channel $\mathcal E_{V|V}$ is a density operator $\rho_{\overline V}$ on $\mathcal H_{\overline V} = \mathcal H_V \otimes \mathcal H_{R_V}$ given by $\rho_{\overline V} = (\mathcal E_{V|V} \otimes \mathcal I_{R_V|R_V})(\kb{\Phi}{\Phi}_{VR_V})$, where $\mathcal I$ denotes the identity channel. For independent error models considered here, $\rho_{\overline V} = \bigotimes_{u \in V} \rho_{\overline u}$.

For each stabilizer generator $S_j$, denote $\overline{S}_j = \bigotimes_{u \in U_j} \sigma^{\alpha_j^u}_u \otimes \sigma^{\alpha_j^u}_{R_u}$, and construct the associated projectors $\overline{P}_j^\pm = (I \pm \overline{S}_j)/2$. An important property of these operator is that they fix the maximally entangled state $\overline{S}_j \Ket{\Phi}_{VR_V} = \overline{P}_j^+ \Ket{\Phi}_{VR_V} = \Ket{\Phi}_{VR_V}$. Let $E$ be an operator on $V$. If $E$ commutes with $S_j$, we have $\overline{P}_j^+ (E\otimes I_{R_V}) \Ket{\Phi}_{VR_V} = (E\otimes I_{R_V}) \Ket{\Phi}_{VR_V}$ and $\overline{P}_j^- (E\otimes I_{R_V}) \Ket{\Phi}_{VR_V} = 0$, while if $E$ anti-commutes with $S_j$, the same identities hold with $\overline P_j^+$ and $\overline P_j^-$ exchanged. It follows from this observation that conditioned on the error syndrome $\bf s$, the channel is described by the Jamio\l kowski matrix
\begin{equation}
\rho_{\overline V|{\bf s}} = \frac 1Z \prod_{j} \overline P_j^{s_j} \po \bigotimes_{v \in V} \rho_{\overline v}, 
\end{equation}
that is a quantum factor graph. 

There are a number of relevant quantities that can be evaluated from this factor graph. For instance, one can efficiently evaluate the conditional channel on any constant size set of qubits $W \subset V$ vial partial trace. This is useful in iterative decoding schemes such as those used for quantum turbo-codes \cite{OPT07a} and low density parity check codes \cite{COT05a}.  In those cases, the conditional channel on $W$ can only be evaluated approximately since it requires loopy QBP. The factor graph also enables exact evaluation of the logical error in a concatenated block coding scheme \cite{Pou06b} such as used in fault-tolerant protocols. 

\subsection{Simulation of Many-Body Quantum Systems}

\label{App:Stat}

In statistical physics, the state of a many-body quantum system $V$ is a Gibbs state $ \rho_V  = \frac 1Z \exp(-\beta H)$ for some Hamiltonian $H$, where $\beta = 1/T$ is the inverse temperature. Typically, $H$ is the sum of single and two-body interactions $H = \sum_{u \in V} H_u + \sum_{(u,w) \in E} H_{uv}$ on some graph $G = (V,E)$. Understanding the correlations present in these states is a great challenge in theoretical physics. In this section, we describe how QBP can serve as an heuristic method to accomplish this task approximately. For an account of the use of Belief Propagation in classical statistical mechanical systems, we refer the reader to the text of M\'ezard and Montanari \cite{MM07a}.

Defining $\mu_u = \exp(-\beta H_u)$ and $\nu_{v:w} = \exp(-\beta H_{vw})$ gives an expression for $\rho_V$ of the form of eq.~\eqref{Graph:QGSEInf}:
\begin{equation}
\rho_V = \left(\bigotimes_{v \in V} \mu_v\right) \odot \left(\bigodot_{(v,w) \in E} \nu_{v:w}\right)
\end{equation}
Thus, $\rho_V$ is an $\infty$-bifactor state. As mentioned in \S\ref{QBP}, a QBP$^{(\infty)}$ algorithm can easily be formulated for this type of bifactor state, and still converge to the exact solutions of the corresponding inference problem when $\rho_V$ is a quantum Markov network and $G$ is a tree. This requires replacing all matrix products $\prod$ by the commutative product $\odot$ in the defining equations of QBP$^{(\infty)}$ eqs.~(\ref{message}-\ref{belief2}). The proof of  convergence Theorem \ref{thm:QBP} under these more general conditions follows essentially the same reasoning.  

To obtain a bifactor state that satisfies the commutation condition $[\nu_{u:v},\nu_{w:x}]=0$, it is possible to coarse-grain $G$ in a way that the resulting interaction between coarse-grained neighbors commute. Consider for instance a one dimensional chain $G = (V,E)$ with $V = \{u\}_{u=1,\ldots,N}$ and $E = \{(u,u+1)\}_{u=1,\ldots,N-1}$. We can construct a coarse-grained graph $\tilde G$ by identifying all vertices $2 u-1$ and $2 u$ for $u=1,\ldots,\lfloor\frac N 2\rfloor$. The state $\rho_V$ is then an $\infty$-bifactor state on $\tilde G$, with operators
\begin{align}
\tilde\mu_u &= \mu_{2u-1} \odot \mu_{2u} \odot \nu_{2u-1:2u} \\
\tilde\nu_{u:u+1} &= \nu_{2u:2u+1},
\end{align}
satisfying $[\tilde{\nu}_{u:u+1},\tilde{\nu}_{v:v+1}]$
Thus, $\infty$-bifactor states are commonplace in quantum many-body physics. Unfortunately, the convergence of the QBP algorithm in this case requires the state to be a quantum Markov network, which cannot be tested directly in general. As we will now explain, it is often possible to reasonably approximate a Gibbs state by an $n$-bifactor with finite $n$, and sometime even $n=1$.

A simple way to obtain an $n$-bifactor state is to approximate $\odot$ by $\pns{n}$ for some large value of $n$. In the context of many-body physics, this is called a Trotter-Suzuki decomposition of the Gibbs state, and becomes more accurate as the ratio $\beta/n$ decreases. The QBP$^{(n)}$ algorithm can then be operated on this $n$-bifactor state, but its convergence again requires some independence condition that cannot be verified systematically. Alternatively, one can use the replica method described in section \ref{Heur:Rep} and solve the inference problem exactly with QBP$^{(1)}$, but with an increase in complexity exponential in $n$. The replica method is then reminiscent of the well known correspondence between quantum statistical mechanics in $d$ dimensions and classical statistical mechanics in $d+1$ dimensions, where the extra dimension represents inverse temperature. 

The 1-bifactor states also capture the correlations of some non-trivial quantum many-body systems. {\em Valence bond solid} (VBS) states were introduced in Ref. \cite{AKLT87a,AKLT88a} as exact ground states (i.e. $T=0$ Gibbs states) of spin systems with interesting properties. Recent work has generalized these constructions to {\em matrix product states} (MPS) in one-dimension \cite{FNW92a,Vid04a,Vid06a}, and {\em projected entangled-pair states} (PEPS) for higher dimensions \cite{VC04a,SDV06a}. These form an important class of states for the description of quantum many-body systems. For instance, {\em density matrix renormalization group} (DMRG) \cite{Whi92a} --- one of the most successful method for the numerical study of spin chains --- is now understood as a variational method over MPS \cite{OR95a,DMNS98a,VPC04b}. All these states are instances of 1-bifactor states.

\begin{figure}[h!]
\center \includegraphics[height=1.6in]{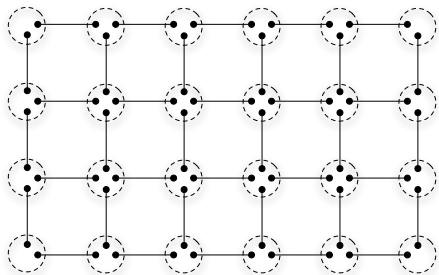}
\caption{Projected entangled pair state on a two-dimensional square lattice. The vertices are associated to dashed circles. Each $\bullet$---$\bullet$ represents a maximally entangled state of $D$ dimension shared between neighboring vertices. A partial isometry $A_u: (\mathbb{C}^D)^{c_u} \rightarrow \mathbb{C}^d$ is applied at each vertex, where $c_u$ is the degree of vertex $u$. }
\label{VBS}
\end{figure}

For sake of simplicity, we will demonstrate this claim for one-dimensional MPS, but the same argument holds for higher dimensions. The MPS $\Ket\Psi$ is a pure state of a collection of $N$ $d$-dimensional quantum systems displayed on a one dimensional lattice. Each vertex $u$ is assigned two ``virtual particles" $L_u$ and $R_u$, where $L$ and $R$ stand for left and right (see Fig.~\ref{VBS} for a illustration of this construction in two-dimensions). Each of these particles are associated a Hilbert space $\mathcal{H}_{L_u} = \mathcal{H}_{R_u} = \mathbb{C}^D$. Initially, the right particle of vertex $u$ is in a maximally entangled state with the left particle of vertex $u+1$; $\Ket{\Phi}_{R_u\cup L_{u+1}} = \frac{1}{\sqrt D} \sum_{\alpha=1}^D \Ket{\alpha}_{R_u}\Ket{\alpha}_{L_{u+1}}$ where $\Ket\alpha$ are orthogonal basis vectors for $\mathbb{C}^D$. (The lattice can be closed to form a circle, in which case we identify $N+1 = 1$.) The initial state is therefore $\Ket{\Phi_0} = \bigotimes_u \Ket{\Phi}_{R_u\cup L_{u+1}}$. 

To obtain the MPS, apply an operator $A_u: \mathcal{H}_{L_u}\otimes \mathcal{H}_{R_u} \rightarrow  \mathbb{C}^d$
\begin{equation}
A_u = \sum_{j=1}^d \sum_{\alpha,\beta = 1}^D A_u^{j,\alpha,\beta} \kb{j}{\alpha,\beta}
\end{equation}
to each vertex of the lattice. The vectors $\Ket j$ form an orthogonal basis for $\mathbb{C}^d$.  The resulting state is
\begin{equation}
\Ket{\Psi} = \bigotimes_{u=1}^N A_u \Ket{\Phi_0} 
\propto \sum_{j_1,j_2,\ldots,j_N=1}^d \Tr{B_1^{j_1}B_2^{j_2}\ldots B_N^{j_N}} \Ket{j_1,j_2,\ldots,j_N}
\label{eq:MPS}
\end{equation}
where the matrices $B_u^j$ are the submatrices of $A_u$ with matrix elements $(B_u^j)_{(\alpha,\beta)} = A_u^{j,\alpha,\beta}$.

For the corresponding 1-bifactor state, the underlying graph $G = (V,E)$ is also a one dimensional lattice $V = \{1,2,\ldots,N\}$ and $E = \{(1,2),(2,3),\ldots ,(N-1,N)\}$.  The Hilbert space associated to vertex $u$ is $\mathcal{H}_u = \mathbb{C}^D \otimes\mathbb{C}^D$. As above, it is convenient to imagine that each vertex $u$ is composed of two $D$-dimensional subsystems $L_u$ and $R_u$. Then, up to a local isometry, the MPS of eq.~\eqref{eq:MPS} can be expressed as a 1-bifactor state eq.~\eqref{Graph:QGSE} with
\begin{equation}
\mu_u = A_u^\dagger A_u \quad \mathrm{and}\quad \nu_{u:v} = \kb{\Phi}{\Phi}_{R_u\cup L_v}.
\end{equation}
Moreover, the operators $\nu_{u:v}$ mutually commute. To see the relation with eq.~\eqref{eq:MPS}, note that the operators $A^u$ can be polar decomposed $A_u = U_u \sqrt{A_u^\dagger A_u} = U_u \mu_u^\frac 12$. \footnote{Note that $\mu_u$ has rank $\leq d$. This can be seen straightforwardly by writing $\mu_u = \sum_{j=1}^d A_u^{*j,\alpha,\beta}A_u^{j,\gamma,\delta} = \sum_{j=1}^d \kb{h_u^j}{h_u^j}$ where $|h_u^j\rangle = \sum_{\alpha,\beta} A_u^{j,\alpha,\beta} \Ket{\alpha,\beta} \in \mathcal{H}_u$.} The matrix $U_u$ is a partial isometry $\mathcal{H}_u \rightarrow \mathbb{C}^d$ and
\begin{align}
\kb\Psi\Psi &=  \frac 1Z \left(\prod_{u \in V} A_u\right) \kb{\Phi_0}{\Phi_0}\left(\prod_{u \in V} A^\dagger_u\right) \\
&= \frac 1Z \left(\prod_{u \in V} U_u\mu_u^\frac 12 \right) \left(\prod_{(v,w) \in E} \nu_{u:v}\right) \left(\prod_{u \in V} \mu_u^\frac 12 U^\dagger_u \right) \\
&= \frac 1Z \left(\prod_{u \in V} U_u \right) \left(\bigotimes_{u \in V} \mu_u \right) \po \left( \prod_{(v,w) \in E} \nu_{u:v}\right) \left(\prod_{u \in V}  U^\dagger_u \right)
\end{align}
as claimed. 

Bifactor states are thus relevant to the description of quantum many-body systems. QBP can sometimes be used to efficiently compute correlation functions, but in general for spatial dimension larger than one, its convergence is not guaranteed. This is mainly due to the presence of small loops in the underlying graph. Partial solutions have been proposed to overcome this difficulty \cite{VC04a}, and it is conceivable that techniques from loopy Belief Propagation and its generalizations \cite{YFW02a} will improve these algorithms. As in the classical case however, QBP may be more appropriate for the study of quantum systems on irregular sparse graphs, such as those encountered in classical spin glasses. 

Finally, it should be noted that the Markov conditions required to certify the convergence of QBP --- or the associated coarse-grained Markov conditions as explained in the previous section --- are weaker than those typically studied in statistical physics, namely the vanishing of connected correlation functions beyond some length scale. For pure quantum states, the two notions coincide and are equivalent to the absence of long-range entanglement.  At finite temperature however, the state is mixed and the vanishing of mutual information between vertices $u$ and $u+\ell$ conditioned on vertices $u+1,\ldots,u+\ell-1$  eq. \eqref{correlation-length} does not imply the absence of connected correlations $\langle A_uA_{u+\ell}\rangle = \Tr{\rho_V A_u A_{u+\ell}} - \Tr{\rho_V A_u}\Tr{\rho_VA_{u+\ell}}$.

\section{Related Work}

\label{Relate}

In this section, our approach to quantum Graphical Models and Belief Propagation is compared to other proposals that have appeared in the literature.  Firstly, Tucci has developed an approach to quantum Bayesian Networks \cite{Tuc95a}, Markov Networks \cite{Tuc07a} and Belief Propagation \cite{Tuc98a} based on a different analogy between quantum theory and classical probability, namely the idea that probabilities should be replaced by complex valued amplitudes.  Tucci's models require that these amplitudes should factorize according to conditions similar to those used in classical Graphical Models. One disadvantage of this is that the definition requires a fixed basis to be chosen for the system at each vertex of the graph, and the factorization condition for Bayesian Networks is not preserved under changes of this basis.  In contrast, our definition of quantum conditional independence is based on an explicitly basis independent quantity, so it does not have this problem.  Another difficulty with using amplitudes is that they are only well-defined for pure states, so that mixed states have to be represented as purifications on larger networks.  In our approach, density operators are taken as primary, so mixed states can be represented without purification.  On the other hand, the Tucci's definitions can easily accommodate unitary time evolution, whereas we do not have a general treatment of dynamics in our approach at the present time.  A related definition of quantum Markov Networks, also based on amplitudes but without a development of the corresponding Belief Propagation algorithm, has been proposed by La Mura and Swiatczak \cite{LMS07a}, to which similar comments apply.

There has also been work on Quantum Markov networks within the quantum probability literature \cite{Lei01a,AF03a,AF03b}, although Belief Propagation has not been investigated in this literature.  This is closer to the spirit of the present work, in the sense that it is based on the generalization of classical probability to a noncommutative, operator-valued probability theory.  These works are primarily concerned with defining the Markov condition in such a way that it can be applied to systems with an infinite number of degrees of freedom, and hence an operator algebraic formalism is used.  This is important for applications to statistical physics because the thermodynamic limit can be formally defined as the limit of an infinite number of systems, but it is not so important for numerical simulations, since these necessarily operate with a finite number of discretized degrees of freedom.  Also conditional independence is defined in a different way via quantum conditional expectations, rather than the approach based on conditional mutual information and conditional density operators used in the present work.  Nevertheless, it seems likely that there are connections to our approach that should to be investigated in future work. 

Lastly, during the final stage of preparation of this manuscript, two related papers have appeared on the physics archive. An article by Laumann, Scardicchio and Sondhi \cite{LSS07a} used a QBP-like to solve quantum models on sparse graphs.  Hastings \cite{Has07b} proposed a QBP algorithm for the simulation of quantum many-body systems based on ideas similar to the ones presented here. The connection between the two approaches, and in particular the application of the Lieb-Robinson bound \cite{LR72a} to conditional mutual information, is worthy of further investigation. 

% ***** 6) CONCLUSIONS *****

\section{Conclusion}

\label{Conc}

\begin{figure}
\center\includegraphics[width=11cm]{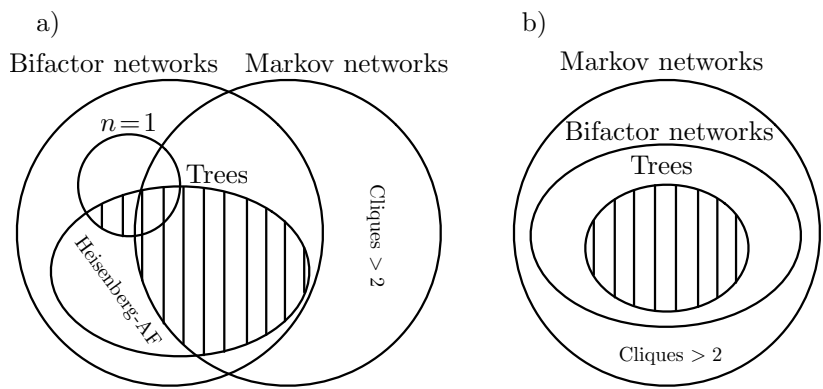}
\caption{Relation between Markov Networks, Bifactor Networks, and 1-Bifactor Networks in a) quantum theory and b) classical probability theory. The hashed regions indicate the domain of convergence of the associated Belief Propagation algorithms. Figure a). Convergence of Belief Propagation on trees for Markov Networks is Theorem \ref{thm:QBP} and for 1-Bifactor states is Corollary \ref{cor:QBP1}. That all Markov Networks on trees are Bifactor states is Theorem \ref{The:mutual}. The existence of Bifactor Networks on trees that are not Markov Networks is given by Example \ref{ex:notMarkov} for $n<\infty$ and the Heisenberg anti-ferromagnetic spin chain of Example \ref{ex:heisenberg} for $n = \infty$. Markov Networks on trees with cliques of size $>2$ are generally not Bifactor Networks, c.f. Theorem \ref{Graph:QHC}. Figure b). That all classical Bifactor Networks are Markov Networks is the Hammersley-Clifford Theorem \ref{thm:HC}, and convergence of Belief Propagation on trees follows from Theorem \ref{thm:QBP}. }
\label{fig:worldview}
\end{figure}

In this paper, we have presented quantum Graphical Models and Belief Propagation based on the idea that quantum theory is a noncommutative, operator-valued, generalization of probability theory.  Our main results are summarized on Fig. \ref{fig:worldview}.  We expect these methods to have significant applications in quantum error correction and the simulation of many-body quantum systems.  We are currently in the process of implementing these algorithm numerically in both of these contexts. Belief Propagation based decoding of several types of quantum error correction codes has already been implemented quite successfully, e.g. on concatenated block codes \cite{Pou06b}, turbo codes \cite{OPT07a}, and  sparse codes \cite{COT05a}. However, for the noise models considered there, the corresponding bifactor states only involve commuting operators and thus the corresponding inference problem could be solved by means of a classical Belief Propagation algorithm.  We conclude with several open questions suggested by this work. 

In the context of many-body physics, it would be interesting to relate the class of solutions obtained by QBP to other approximation schemes used in statistical physics, much in the spirit of the work of Yedidia \cite{Yed01a} in the classical setting. A related problem would be to understand how the different classes of bifactor states relate to each other. We suspect that when the Hilbert space dimension at each vertex of the graph is held fixed, the $n$-bifactor states on that graph form a subset of the $m$-bifactor states when $n<m$. If that conjecture were true, it might lead to a family of approximation schemes converging to the correct solution. It would also reveal an interesting discrepancy between the classical and quantum settings. Classically, the problem of computing correlation functions in a disordered many-body system and the problem of decoding an error correction code are equivalent. If our conjecture holds true, in the quantum case the latter is simpler than the former.

Whilst our definition of a quantum Markov Network is well motivated as a direct analog of a classical Markov Network, it does not seem to represent the most general class of states to which our Belief Propagation algorithms are applicable.  In particular, in \S\ref{QBP:Conv} it was shown that QBP converges on trees for arbitrary bifactor states defined with respect to the $\po$ product.  One reason for this discrepancy might be that the quantum conditional independence condition, $I_\rho (U,W|X)$, only allows classical correlations to be mediated between $U$ and $W$ via $X$, i.e. $\rho_{U\cup W}$ is always separable, whereas the classical condition $I_P(U,W|X)$ is compatible with an arbitrary distribution $P(U \cup W)$.  This suggests that quantum conditional independence could be relaxed to a condition that allows quantum correlations, i.e. entanglement, to be mediated by $X$, whilst still preserving the validity of Belief Propagation.  It would be interesting to find a condition like this that also satisfies the graphoid axioms, so that it could naturally be represented on a graph.

Nevertheless, quite apart from their application in Belief Propagation algorithms, the mathematical structures investigated in this work should be of interest in other areas of quantum information and computation.  Firstly, the characterizations of quantum conditional independence in terms of conditional density operators given in \S\ref{Cond:CDO} should be useful, and indeed are currently being applied to the problem of pooling quantum states \cite{LS07a}.  Another interesting area of investigation would be the computational complexity of inference on quantum Markov Networks.  In the classical case, it is fairly straightforward to find families of Markov Networks that encode instances of NP complete problems, such as satisfiability or graph colorability.  Therefore, one would expect to be able to encode problems that are similarly hard for quantum computers, i.e. complete for the complexity class QMA, as inference problems on quantum Markov Networks.  This should be closely related to the quantum marginals problem, which has recently be proved to be QMA-complete \cite{AGK07a,Ira07a}.

Finally, this work leaves open the question of fully characterizing quantum Markov Networks.  The most generally applicable result given here is theorem \ref{Graph:QHC}, which is a direct analog of one direction of the classical Hammersely Clifford theorem using the $\pity$ product.  A full characterization would provide a converse to this theorem, i.e. a set of conditions on the operators in eq.~\eqref{Graph:QHCDecomp}, satisfied by the construction used in the proof, such that all states of this form are guaranteed to satisfy the Markov condition.  Analogous theorems for the $\pns{n}$ products would also be useful. This work also leaves open the question of intersection for quantum conditional mutual information, i.e. whether $S(U:W|X\cup Y) = 0$ and $S(U:Y|W\cup X) = 0$ imply $S(U:W\cup Y|X) = 0$ for strictly positive states. This result would imply that positive quantum Markov networks obey global Markov properties.  

\section*{Acknowledgments}

ML would like to thank Rob Spekkens for useful discussions about quantum conditional independence. DP is grateful to Harold Ollivier for many stimulating discussion on Belief Propagation.   

At IQC, ML was supported in part by MITACS and ORDCF.  Research at Perimeter Institute for Theoretical Physics is supported in part by the Government of Canada through NSERC and by the Province of Ontario through MRI.  ML was also supported in part by grant RFP1-06-006 from The Foundational Questions Institute (fqxi.org). 

DP is supported in part by the Gordon and Betty Moore Foundation through Caltech's Center for the Physics of Information, by the National Science Foundation under Grant No. PHY-0456720, and by the Natural Sciences and Engineering Research Council of Canada.

% ***** APPENDICES *****

\appendix
\section[Useful Notation]{A useful notation for probability distributions and density matrices}

\label{ProbNot}

\subsection{Probability Distributions}

In standard Kolmogorov probability theory for finite sample spaces, probabilities are given by a measure $\mu$ on a sample space $(\Omega, 2^\Omega)$, where $\Omega$ is a set of elementary events and $2^\Omega$ is the power set, i.e. the set of all subsets of $\Omega$.  Specifically, $\mu: 2^\Omega \rightarrow [0,1]$ and satisfies the axioms
\begin{align}
& \forall \Lambda \in 2^\Omega, \quad 0 \leq \mu(\Lambda) \leq 1\\
& \mu(\Omega) = 1\\
& \text{If} \,\, \Lambda_1, \Lambda_2, \ldots \Lambda_d \,\, \text{are disjoint sets in} \,\, 2^\Omega \,\, \text{then} \,\, \mu(\cup_{j=1}^d \Lambda_j) = \sum_{j=1}^d \mu(\Lambda_j).
\end{align}
In particular, this implies that $\mu(\emptyset) = 0$ and $\forall \Lambda_1, \Lambda_2 \in 2^\Omega$,
\begin{align}
\mu(\Lambda_1 \cup \Lambda_2) \geq & \mu(\Lambda_1) \\
\mu(\Lambda_1 \cap \Lambda_2) \leq & \mu(\Lambda_1) \\
\text{If}\,\, \Lambda_1 \subseteq \Lambda_2 \,\, \text{then} \,\, \mu(\Lambda_1) \leq & \mu(\Lambda_2).
\end{align}
The conditional probability of $\Lambda_2$, given $\Lambda_1$ is defined to be
\begin{equation}
\text{Prob}(\Lambda_2|\Lambda_1) = \frac{\mu(\Lambda_1 \cap \Lambda_2)}{\mu(\Lambda_1)}
\end{equation}
provided $\mu(\Lambda_1) \neq 0$ and is undefined otherwise.  In particular, for any $\Lambda \in 2^{\Omega}$, this means that $\text{Prob}(\Lambda|\emptyset)$ is always undefined and that $\text{Prob}(\Lambda|\Omega) = \mu(\Lambda)$.

Our notation for probability distributions over random variables works in an almost exactly opposite way to the Kolmogorov conventions, but is very convenient for the discussion of Graphical Models.  For a random variable $v$ that takes a finite number of possible values, write $P(v)$ for the probability distribution of $v$. For definiteness, suppose that $v$ takes integer values $\{1,2,\ldots d\}$.  Then, a sample space can be associated with $v$ by setting $\Omega_v = \left \{ v =1, v=2, \ldots, v=n\right \}$, and a measure $\mu:2^{\Omega_v} \rightarrow [0,1]$ can be defined on this space.  The notation $P(v)$ is a stand in for $\mu(v = j)$ when $j$ is an arbitrary unspecified value.  To give some precise examples of how this works, let $f$ be a function with domain $\{ 1,2, \ldots, d\}$ and let $g$ be a function with domain $[0,1]$.  Then, the expression $g(P(v)) = f(v)$ is interpreted as $\forall j, \, g(\mu(v=j)) = f(j)$, and the expression $\sum_{v}g(P(v))f(v)$ is interpreted as $\sum_j g(\mu(v=j))f(j)$.  It is straightforward to see how this generalizes to more complicated examples.

Now consider the case of two random variables $v,w$ for which we can set up sample spaces $\Omega_v$ and $\Omega_w$ as above.  Joint probabilities are given by a measure $\mu$ on the sample space $\left ( \Omega_v \times \Omega_w, 2^{\Omega_v \times \Omega_w} \right )$.  The notation $P(v,w)$ stands for $\mu \left (v = j \times w = k \right )$, where both $j$ and $k$ are arbitrary unspecified values.  Note that
\begin{equation}
\mu \left (v = j \times w = k \right ) = \mu \left ( \left ( v = j \times \Omega_w \right ) \cap \left ( \Omega_v \times w = k \right ) \right ).
\end{equation}

The notation P(v,w) can be made precise in the same way as the examples given above for a single variable, but two additional definitions are worthy of note.  Firstly, the marginal probability of $v$ is written as $P(v) = \sum_w P(v,w)$ and this corresponds to the equation
\begin{equation}
\mu(v = j \times \Omega_w) = \sum_k \mu(v = j \times w = k).
\end{equation}
Secondly, the conditional probability of $w$ given $v$ is written as $P(w|v) = \frac{P(v,w)}{P(v)}$, which corresponds to
\begin{equation}
\begin{split}
\text{Prob} \left ( \Omega_v \times w = k | v = j \times \Omega_w \right ) & = \frac{\mu \left ( \left ( v = j \times \Omega_w \right ) \cap \left ( \Omega_v \times w = k \right ) \right )}{\mu(v = j \times \Omega_w)} \\ & = \frac{\mu(v = j,w = k)}{\mu(v = j \times \Omega_w)}.
\end{split}
\end{equation}
The generalization of this to arbitrary numbers of random variables is straightforward.

The present notation can be extended to a set of random variables $V = \left \{ v_1, v_2, \ldots, v_N \right \}$, where $v_j$ is a random variable taking values in $\{1,2,\ldots, d_j\}$.  Consider the joint probability distribution of an arbitrary subset $U \subseteq V$.  Let $I = \{ i_1, i_2, \ldots, i_M\}$ be the index set of $U$, i.e. the subset of $\{1,2, \ldots, N\}$ consisting of the indices of the $v_j$'s that are contained in $U$.  Then define $P(U) = P(v_{i_1},v_{i_2}, \ldots, v_{i_M})$.  This implies that $P(\emptyset) = 1$, which is opposite to the Kolmogorov convention for events, but recall that here $\emptyset$ is an empty set of random variables rather than an event in a sample space.  To see this, note that the expression $P(U)$ may be read as meaning that the variables in $U$ are constrained to take particular values, whilst the variables in $V-U$, the relative complement of $V$ in $V$, may take any value.  Thus $P(\emptyset)$ is the probability of the event corresponding to no constraints, i.e. the entire sample space.   More precisely, if we define$K = \{k_1,k_2,\ldots,k_{N-M}\}$ to be the index set of $V-U$ and let $j_{1},j_{2}, \ldots, j_{M}$ be particular instantiations of $v_{i_1},v_{i_2}, \ldots, v_{i_M}$, then $P(U)$ corresponds to $\mu(v_{i_1} = j_1 \times v_{i_2} = j_2 \times \ldots \times v_{i_M} = j_M \times \Omega_{v_{k_1}} \times \Omega_{v_{k_2}} \times \ldots \times \Omega_{k_{N-M}})$.  Thus, for $U = \emptyset$ we have $P(\emptyset) = \mu(\Omega_{v_1} \times \Omega_{v_2} \times \ldots \times \Omega_{v_N}) = 1$ via the standard Kolmogorov axioms.

All the usual set theoretic notions can be applied at the level of random variables, and it is straightforward to verify that the following relations hold for all $U, W \subseteq V$
\begin{align}
& P(U \cup W) \leq P(U) \\
& P(U \cap W) \geq P(U) \\
& \text{If} \,\, U \subseteq W \,\, \text{then} \,\, P(U) \leq P(W).
\end{align}
Conditional probabilities $P(W|U)$ are only well-defined for disjoint subsets, so $P(W|V)$ is always undefined and $P(W|\emptyset) = P(W)$.

Finally, note that this notation introduces an ambiguity for singleton sets $\{v\}$, since $P(v)$ and $P(\{v\})$ denote the same object.  These are used interchangeably and set theoretic operations like $U \cup \{v\}$ are denoted $U \cup v$ when this does not cause ambiguity.

\subsection{Density Matrices}

For quantum theory, the corresponding notation is obtained by replacing random variables $v$ with  finite-dimensional Hilbert spaces $\mathcal{H}_v$ and $P(v)$ with a density matrix $\rho_v$ acting on $\mathcal{H}_v$.  The density matrix $\rho_v$ is referred to as the state of system $v$, with the fact that it is defined on a corresponding Hilbert space $\mathcal{H}_v$ left implicit.  If we have a set $V$ of $N$ quantum systems $V = \{v_1,v_2, \ldots, v_N\}$, then the state $\rho_V$ is defined on the Hilbert space $\mathcal{H}_{v_1} \otimes \mathcal{H}_{v_2} \otimes \ldots \otimes \mathcal{H}_{v_N}$.  For an arbitrary subset $U \subseteq V$, the state $\rho_U$ is defined to be the partial trace of $\rho_V$ over all the systems in $V - U$.  With this convention, $\emptyset$ is associated with the trivial Hilbert space $\mathbb{C}$, so that $\rho_{\emptyset} = 1$.  It is convenient to suppress tensor products with identity operators in order to equate operators acting on different subsets of $V$.  Explicitly, if $U,W \subseteq V$ and $A_U$ and $B_W$ are operators acting on $\mathcal{H}_U$ and $\mathcal{H}_W$ respectively then $A_U = B_W$ is defined to mean $A_U \otimes I_{W - (U \cap W)} = B_W \otimes I_{U- (U \cap W)}$.  Generally, identity operators are omitted in this way unless their presence is required to clarify an argument.

\section{Proof of Theorem \ref{Graph:QHC}}

\label{Proof}

\begin{Lem}
\label{Mobius}
Let $V$ be a collection of quantum systems with Hilbert space $\mathcal{H}_V = \bigotimes_{v \in V} \mathcal{H}_v$ and let $H_V$ be an operator on $\mathcal{H}_V$.  Let $\Ket{\alpha}_v \in \mathcal{H}_v$ be a set of pure states, where $\Ket{\alpha}_v$ may be a different state for each $v$, and for $U \subseteq V$ define $\Ket{\alpha}_U = \bigotimes_{v \in U} \Ket{\alpha_v}$.  For all $U \subseteq V$ define
\begin{equation}
J_U = \Bra{\alpha}_{V - U} H_V \Ket{\alpha}_{V - U} \otimes I_{V - U},
\end{equation}
where $V - U$ denotes the relative complement of $U$ in $V$, and 
\begin{equation}
\label{TheKs}
K_U = \sum_{W \subseteq U} (-1)^{\left | U - W \right |} J_W,
\end{equation}
where $| \cdot |$ denotes the order, i.e. number of elements contained in, a set.  Then,
\begin{equation}
\label{MobiusRes}
H_V = \sum_{U \subseteq V} K_U.
\end{equation}
\end{Lem}

\begin{proof}
Consider the double sum expression obtained by substituting eq.~\eqref{TheKs} into the right hand side of eq.~\eqref{MobiusRes}.
\begin{equation}
\label{DoubleSum}
\sum_{U \subseteq V} \sum_{W \subseteq U}  (-1)^{\left | U - W \right |} J_W.
\end{equation}
Note that the coefficient of $J_W$ in this expression is
\begin{equation}
\label{DoubleSum2}
\sum_{\{ U : W \subseteq U \subseteq V \}} (-1)^{|U - W|} = \sum_{X \subseteq (V - W)} (-1)^{|X|}.
\end{equation}
If $W = V$ then $\emptyset$ is the only subset of $V - W$, so the last sum reduces to $(-1)^0 = 1$.  The corresponding term in eq.~\eqref{DoubleSum} is just $H_V$, so it just remains to prove that all the other terms sum to $0$.  For $W \neq V$, choose an arbitrary element $v \in (V - W)$.  Let $\mathfrak{X} = \left \{ X \subseteq (V - W)| v \notin X \right \}$ and let $\tilde{\mathfrak{X}} = \left \{ X \subseteq (V - W)| v \in X \right \}$.  For each $X \in \mathfrak{X}$, define $\tilde{X} \in \tilde{\mathfrak{X}}$ via $\tilde{X} = X \cup \{v\}$.  This correspondence is a bijection, so exactly half of the subsets of $V - W$ contain $v$ and the other half do not contain $v$.  Further, if $X \in \mathfrak{X}$ has even order then $\tilde{X}$ has odd order, and if $X \in \mathfrak{X}$ has odd order then $\tilde{X}$ has even order.  Thus, there are an equal number of odd and even order subsets of $V - W$, so the right hand side of eq.~\eqref{DoubleSum2} is zero.
\end{proof}

\begin{Lem}
\label{ZeroLemma}
Let $V$ be a collection of quantum systems with Hilbert space $\mathcal{H}_V = \bigotimes_{v \in V} \mathcal{H}_v$ and let $H_V$ be an operator on $\mathcal{H}_V$.  Let $\Ket{\alpha}_v \in \mathcal{H}_v$ be a set of pure states.  For nonempty $U \subseteq V$ define $K_U$ as in eq.~\eqref{TheKs} and let $u \in U$.  Then
\begin{equation}
\Bra{\alpha}_u K_U \Ket{\alpha}_u = 0
\end{equation}  
\end{Lem}
\begin{proof}
Let $W \subseteq U$.  If $u \notin W$ then $\Bra{\alpha}_{u} J_W \Ket{\alpha}_u = \Bra{\alpha}_{V - W} H_V \Ket{\alpha}_{V - W} I_{V - (W \cup \{u\})}$.  Also,
\begin{eqnarray}
\Bra{\alpha}_{u} J_{W \cup \{u\}} \Ket{\alpha}_u & = & \Bra{\alpha}_{V - (W \cup \{u\})} \Bra{\alpha}_u H_V \Ket{\alpha}_{V - (W \cup \{u\})} \Ket{\alpha}_u I_{V - (W \cup \{u\})} \\
& = & \Bra{\alpha}_{V - W} H_V \Ket{\alpha}_{V - W} \otimes I_{V - (W \cup \{u\})} \\
& = & \Bra{\alpha}_u J_W \Ket{\alpha}_u.
\end{eqnarray}
From the same argument that was used in lemma \ref{Mobius}, the element $u$ divides the subsets of $U$ into pairs, i.e. those that don't contain $u$ and those obtained by adding $u$ to such a set.  As shown above, the operator obtained by projecting the $J$ operator onto $\Ket{\alpha}_u$ is the same for each such pair of subsets, but they enter into eq.~\eqref{TheKs} with opposite sign and so the corresponding terms in $\Bra{\alpha}_u K_U \Ket{\alpha}_u$ cancel.
\end{proof}

\begin{proof}[Proof of Theorem \ref{Graph:QHC}]
Apply lemma \ref{Mobius} with $H_V = \log \rho_V$ and set $\sigma_U = \exp(K_U)$ for all $U \subseteq V$.  Rewriting eq.~\eqref{MobiusRes} in terms of these operators gives
\begin{equation}
\rho_{V} = \odot_{U \subseteq V} \sigma_{U}.
\end{equation}
It remains to show that $\sigma_U$ is the identity whenever $U \notin \mathfrak{C}$, which is equivalent to proving that $K_U = 0$.

For any $U \notin \mathfrak{C}$, we can find two vertices $u,t \in U$ that are not connected by an edge.  In particular, this means that $t \notin n(u)$. Then, the Markov condition, $I \left (\{u\}:V - \left ( \{u\} \cup - n(u) \right ) |n(u) \right )$, implies that
\begin{eqnarray}
\log \rho_{u|V - \{u\}} & = & \log \rho_{u|n(u)} \otimes I_{V - (\{u\} \cup n(u))} \\
& = & \log \rho_{u|n(u)} \otimes I_{V - (\{u\} \cup \{t\} \cup n(u))} \otimes I_t \label{MarkovDecomp}.
\end{eqnarray}
Now, let $\mathfrak{U} = \{ W \subseteq U| u \notin W\}$ and let $\tilde{\mathfrak{U}} = \{ W \subseteq U| u \in W \}$.  As before, every $W \in \mathfrak{U}$ is in one-to-one correspondence with a $\tilde{W} \in \tilde{\mathfrak{U}}$ defined by $\tilde{W} = W \cup \{u\}$, and so eq.~\eqref{TheKs} may be rewritten as
\begin{equation}
K_U = \sum_{W \in \mathfrak{U}} (-1)^{|U - W|} (J_W - J_{\tilde{W}}).
\end{equation}
Next, consider a particular $W$ and the corresponding term $J_W - J_{\tilde{W}}$.  Using the standard rules of conditional density operators,
\begin{eqnarray}
J_W & = & \Bra{\alpha}_{V-W} \log \rho_V \Ket{\alpha}_{V - W} \otimes I_{V-W} \\
& = & \Bra{\alpha}_{V - W} \log \rho_{u|V - u} \Ket{\alpha}_{V - W} \otimes I_{V - W} \nonumber \\
&& + \Bra{\alpha}_{V - W} \log \rho_{V - u} \otimes I_u \Ket{\alpha}_{V - W} \otimes I_{V - W}  \\
& = & \Bra{\alpha}_{V - W} \log \rho_{u|V - u} \Ket{\alpha}_{V - W} \otimes I_{V - W} \label{ToughW} \nonumber\\ 
& & + \Bra{\alpha}_{V - (W \cup \{u\})} \log \rho_{V - u} \Ket{\alpha}_{V (W \cup \{u\})} \otimes I_{V - (W \cup \{u\})} \otimes I_{u}. \label{EasyW}
\end{eqnarray}
Similarly, $J_{\tilde{W}}$ may be written as
\begin{eqnarray}
J_{\tilde{W}} & = & \Bra{\alpha}_{V-\tilde{W}} \log \rho_V \Ket{\alpha}_{V - \tilde{W}} \otimes I_{V-\tilde{W}} \\
& = & \Bra{\alpha}_{V- (W \cup \{u\})} \log \rho_V \Ket{\alpha}_{V - (W \cup \{u\})} \otimes I_{V- (W \cup \{u\})} \\
& = & \Bra{\alpha}_{V - (W \cup \{u\})} \log \rho_{u|V - u} \Ket{\alpha}_{V - (W \cup \{ u\})} \otimes I_{V - (W \cup \{u\})} \nonumber \\
& & + \Bra{\alpha}_{V - (W \cup \{ u\})} \log \rho_{V - u} \otimes I_u \Ket{\alpha}_{V - (W \cup \{u\})} \otimes I_{V - (W \cup \{u\})}  \\
& = & \Bra{\alpha}_{V - (W \cup \{u\})} \log \rho_{u|V - u} \Ket{\alpha}_{V - (W \cup \{u\})} \otimes I_{V - (W \cup \{u\})} \label{ToughtW} \nonumber \\
& & + \Bra{\alpha}_{V - (W \cup \{u\})} \log \rho_{V - u} \Ket{\alpha}_{V - (W \cup \{u\})} \otimes I_{V - (W \cup \{u\})} \otimes I_{u}. \label{EasytW}
\end{eqnarray}
The last terms \eqref{EasyW} and \eqref{EasytW} are identical, so they cancel in $J_W - J_{\tilde{W}}$.  Therefore, $J_W - J_{\tilde{W}}$ is just the difference of \eqref{ToughW} and \eqref{ToughtW}.  The remainder of the proof show that $\Bra{\alpha}_t J_W - J_{\tilde{W}} \Ket{\alpha}_t \otimes I_t = J_W - J_{\tilde{W}}$.  From this it follows that $\Bra{\alpha}_t K_U \Ket{\alpha}_t \otimes I_t = K_U$, but lemma \ref{ZeroLemma} shows that $\Bra{\alpha}_t K_U \Ket{\alpha}_t = 0$, so this is enough to complete the proof.

There are two cases to deal with, either $t \notin W$ or $t \in W$.  When $t \notin W$, both $V - W$ and $V - (W \cup \{u\})$ contain $t$.  The effect of projecting out $\Ket{\alpha}_t$ on terms \eqref{ToughW} and \eqref{ToughtW} is to replace $I_{V - W}$ and $I_{V - (W \cup \{u\})}$ with $I_{V - (W \cup \{t\})}$ and $I_{V - (W \cup \{u\} \cup \{t\})}$ respectively, but then tensoring with $I_t$ restores the original identity operator so both terms are unaffected.  In the case where $t \in W$, we make use of the Markov condition in the form of eq.~\eqref{MarkovDecomp}.  The important point is that $\rho_{u|V - u}$ is of the form $\tau_{V - t} \otimes I_t$, so projecting out $\Ket{\alpha_t}$ and retensoring with $I_t$ again has no effect on the terms \eqref{ToughW} and \eqref{ToughtW}.
\end{proof}

%\bibliographystyle{/Users/dpoulin/archive/qubib}
%\bibliography{/Users/dpoulin/archive/qubib}

\end{document}